\documentclass[12pt]{article}
\usepackage[latin1]{inputenc}
\usepackage{amsfonts,amssymb,amsmath, epsfig}
\usepackage{color,graphicx,graphics}
\usepackage{amsthm}
\usepackage{amsmath,amstext,amssymb,amsfonts, amscd}
\usepackage[lofdepth,lotdepth]{subfig}
\usepackage{xcolor}
\usepackage{hyperref}          
\usepackage{cancel}
\usepackage{stix}

\def\Box{\leavevmode\vbox{\hrule
     \hbox{\vrule\kern4pt\vbox{\kern4pt}%
           \vrule}\hrule}}
\def\blackbox{\leavevmode\vrule height 5pt width 4pt depth 0pt\relax}
\def\endproof{\null\hfill {$\blackbox$}\bigskip}

\def\paragraph#1{{\bf #1\ }}

\newtheorem{lemma}{Lemma}[section]  

\newtheorem{theorem}[lemma]{Theorem}

\newtheorem{corollary}[lemma]{Corollary}

\newtheorem{proposition}[lemma]{Proposition}

\newtheorem{remark}{Remark}[section]

\newtheorem{hypothesis}{Hypothesis}[section]

\title{Binary particle collisions with mass exchange} 
\author{Pierre Degond$^{(1)}$, Jian-Guo Liu$^{(2)}$} 
\date{} 
\begin{document}

\maketitle

\vspace{0.5 cm}

\begin{center}
(1) Institut de Math\'ematiques de Toulouse ; UMR5219 \\
Universit\'e de Toulouse ; CNRS \\
UPS, F-31062 Toulouse Cedex 9, France \\
email: pierre.degond@math.univ-toulouse.fr
\end{center}

\begin{center}
(2) Department of Physics and Department of Mathematics, \\
Duke University, Durham, NC27708, USA \\
email: jliu@phy.duke.edu
\end{center}

\vspace{0.5 cm}
\begin{abstract}
We investigate a kinetic model for interacting particles whose masses are integer multiples of an elementary mass. These particles undergo binary collisions which preserve momentum and energy but during which some number of elementary masses can be exchanged between the particles. We derive a Boltzmann collision operator for such collisions and study its conservation properties. Under some adequate assumptions on the collision rates, we show that it satisfies a H-theorem and exhibit its equilibria. We formally derive the system of fluid equations that arises from the hydrodynamic limit of this Boltzmann equation. We compute the viscous corrections to the leading order hydrodynamic equations on a simplified collision operator of BGK type. We show that this diffusive system can be put in the formalism of nonequilibrium thermodynamics. In particular, it satisfies Onsager's reciprocity relation and entropy decay. 
\end{abstract}

\medskip
\noindent
{\bf Acknowledgements:} PD holds a visiting professor association with the Department of Mathematics, Imperial College London, UK.


\vskip 0.4cm

\setcounter{equation}{0}
\section{Introduction}
\label{intro}

This paper is devoted to the study of a system of interacting particles whose masses are integer multiples of an elementary mass. These particles undergo binary collisions which preserve momentum and energy but during which some number of elementary masses can be exchanged between the particles. So, this system combines the features of a coagulation-fragmentation problem and of a classical rarefied gas dynamics system. 

Coagulation-fragmentation equations describe the size evolution of particle clusters when such clusters can merge or split. There are many different models of such phenomena depending on how many clusters or what cluster sizes are involved in a merge or split event, or whether the cluster sizes are described by discrete or continuous variables. The emblematic example of such a model is the so-called Smoluchowski equation. There is an abundant literature on the mathematical analysis of this equation and its variants, in particular on its dynamics and its equilibria, which will be our main interest here. We refer e.g. to 
\cite{AB1979,C-JSP2007,Carr1992,CDC1994,degond2017coagulation,FM2004,LM2003,LR2015} as well as the short review \cite{wattis2006introduction} and the book~\cite{banasiak2019analytic} for further references. Coagulation-fragmentation equations have many applications in aerosol dynamics \cite{ferreira2021coagulation}, emulsion polymerization \cite{cheng2018particle}, combustion \cite{kraft2005modelling, raj2009statistical} and animal group-size statistics \cite{degond2017coagulation,Ma_etal_JTB11}. 

Most of the literature deals with spatially homogeneous systems, i.e., systems where the size distribution of the clusters do not depend on position. Some models though have considered spatially inhomogeneous situations. In \cite{carrillo2009rigorous}, the coagulation-fragmentation dynamics is coupled with a spatial diffusion. \cite{guiacs2009stochastic,lee2015stochastic} consider spatially discrete versions of the transport processes applying to the clusters. In \cite{goudon2013fluid}, the clusters are transported by a background fluid and modelled by a kinetic equation of Fokker-Planck type. The authors study hydrodynamic limits which roughly lead to the model of \cite{carrillo2009rigorous}, coupled with the movement of the fluid. In \cite{jabin2006kinetic}, a spatially inhomogeneous kinetic fragmentation model is considered and existence of solutions is proved. 

In the present paper, we derive a Boltzmann collision operator for particles which have integer masses and which undergo binary collisions preserving the total mass (i.e. the sum of the masses of the incoming particles). However, the way the total mass is distributed among the two particles may be different in the incoming and outgoing particle pairs. We also assume that the binary collisions preserve the total momentum and total energy. For this reason, we exclude the case of complete coagulation (i.e. there is only one outgoing particle) or complete breakup (i.e. there is only one incoming particle), because in such case, it is impossible to conserve both the total momentum and total energy. We refer to processes in which two particles exchange mass (without merge or breakup) during a collision as a mass-exchange process.  

Collisions with mass exchange may occur for instance in reacting gases \cite{bisi2023internal}, sprays \cite{anidjar1995mass,post2002modeling,rigopoulos2010population,roquejoffre2001kinetic} or heavy ion collisions \cite{ngo1977mass}. However, in the first case, the masses span a finite set of integers and, additionally, the collisions are not kinematically elastic as there are potential energy barriers that the particles must overcome to form reaction products. In the second and third cases, the particles have some internal energy, reflecting vibrating modes in droplets or in ions. Moreover, in the case of sprays, most often passive advection by the background fluid is the main driver of motion.

Here, we construct a simpler model, aiming that both the collision kinetics and mass exchange processes have similar magnitudes and contribute in a comparable way to motion, particle statistics, equilibria and ensemble dynamics. So, we assume that the mass exchange process does not require any potential energy nor does it trigger any internal degrees of freedom of the particles. These assumptions could be satisfied at leading order when such potential energies or internal degrees of freedom are small enough to be treated as perturbations. The modelling is based on the construction of a Boltzmann collision operator akin to that used in rarefied gas dynamics, but which incorporates the mass-exchange phenomenon. The Boltzmann equation of rarefied gas dynamics has been the subject of numerous mathematical works, see \cite{cercignani2013mathematical,saint2009hydrodynamic,ukai2006mathematical,villani2002review} for reviews. 

Another instance which motivated this study is the description of animal groups movements. Animal group size statistics has been treated so far in a purely spatially homogeneous way \cite{degond2017coagulation,Ma_etal_JTB11}. However, one may be interested in investigating to what extent group merges and splits influence the displacement of animals during migrations or foraging. Obviously, momentum and energy are not conserved during animal movements due to self-propulsion. However, interaction mechanisms between animals and their collective dynamics are complex, diverse and still poorly known \cite{vicsek2012collective}. So, rarefied gas binary collisions appear as a starting point for the investigation of mass exchange processes. The study of more complex models with gradually more realistic assumptions will be deferred to future works. 

The outline of this work is as follows. In Section \ref{sec_model}, we derive the Boltzmann operator for mass exchange collisions (referred to as the BME operator) and study its properties (conservations, entropy dissipation, equilibria, etc.). In Section \ref{sec_Boltz_eq}, we investigate the dynamics of the BME equation at large spatio-temporal scales compared with the typical collision scales. This regime is characterized by a small parameter $\varepsilon$ (known as the Knudsen number) and the limit $\varepsilon \to 0$ is called the hydrodynamic limit. In this limit, the system is described by the local mass, momentum and energy densities, as well as the local particle number density (here, since the particles have different masses, the mass and number densities are not proportional). The resulting system will be referred to as the Euler system with mass exchange (EME). Then, we investigate the ${\mathcal O}(\varepsilon)$ correction to the EME system using the Chapman-Enskog procedure. Because computations may be quite complex with the original BME operator, we use a simpler BGK-type collision operator for this study. This leads to a system called the Navier-Stokes equations for mass exchange (NSME). Finally, Section \ref{sec_entropy} is devoted to investigating the entropy properties of the EME and NSME systems. In particular, we show that the NSME system is consistent with the formalism of nonequilibrium thermodynamics \cite{de2013non}: it satisfies Onsager's reciprocity relation and entropy dissipation, and is therefore compatible with the second principle of thermodynamics. Finally, we end this work by a conclusion in Section \ref{sec_conclu}. A short appendix recalls some basic formulas used throughout this work. 

\setcounter{equation}{0}
\section{Boltzmann operator for mass exchange collisions}
\label{sec_model}

\subsection{The collision rule }
\label{subsec_coll_rule}

We first consider two particles of masses $m$ and $m_1$ and velocities $v$ and $v_1$ respectively which interact and give rise to particles with masses $m'$ and $m'_1$ and velocities $v'$ and $v'_1$ respectively. Here, we assume that masses are integers in $\{1, 2, \ldots, \}$ and the velocities are vectors in ${\mathbb R}^n$. The collision obviously preserves the number of particles. We assume that the collision also preserves mass, momentum and energy, i.e. we have 
\begin{eqnarray}
&& m+m_1 = m' + m'_1, \label{eq:cons_mass}\\
&& m v + m_1 v_1 = m' v' + m'_1 v'_1, \label{eq:cons_momentum}\\
&& m |v|^2 + m_1 |v_1|^2 = m' |v'|^2 + m'_1 |v'_1|^2. \label{eq:cons_energy}
\end{eqnarray}
Since zero masses are not allowed, we have $m'$, $m'_1 \in \{1, 2, \ldots, m+m_1-1\}$. We introduce the center-of-mass velocity $v_{\textrm{CM}}$. Because of \eqref{eq:cons_mass} and \eqref{eq:cons_momentum}, we have 
\begin{equation}
v_{\textrm{CM}} = \frac{m v + m_1 v_1}{m+m_1} = \frac{m' v' + m'_1 v'_1}{m'+m'_1}. 
\label{eq:COM_vel}
\end{equation}
We then introduce the velocities in the center-of-mass frame 
\begin{eqnarray}
u &=& v-v_{\textrm{CM}} = \frac{m_1}{m+m_1} (v-v_1), \label{eq:express_u}\\
u_1 &=& v_1-v_{\textrm{CM}} = - \frac{m}{m+m_1} (v-v_1), \label{eq:express_u1}\\
u' &=& v'-v_{\textrm{CM}} = \frac{m'_1}{m'+m'_1} (v'-v'_1), \label{eq:express_u'}\\
u'_1 &=& v'_1-v_{\textrm{CM}} = - \frac{m'}{m'+m'_1} (v'-v'_1). \label{eq:express_u1'}
\end{eqnarray}
Thanks to \eqref{eq:cons_mass} and \eqref{eq:cons_momentum}, Eq. \eqref{eq:cons_energy} can be written
$$ m |u|^2 + m_1 |u_1|^2 = m' |u'|^2 + m'_1 |u'_1|^2, $$
and with \eqref{eq:express_u} to \eqref{eq:express_u1'}, we find
\begin{equation}
(m m_1)^{1/2} |v-v_1| = (m' m'_1)^{1/2} |v'-v'_1|. 
\label{eq:reduced_ener}
\end{equation}
Following the case of equal masses, we can write 
\begin{equation} 
(m' m'_1)^{1/2} (v'-v'_1) = (m m_1)^{1/2} \big[ v-v_1 - 2 (v-v_1) \cdot \Omega  \, \Omega \big], 
\label{eq:v'-v1'}
\end{equation}
where $\Omega \in {\mathbb S}^{n-1}$ with $\Omega \cdot (v-v_1) \leq 0$. Inserting it into \eqref{eq:express_u'} and \eqref{eq:express_u1'}, we get
\begin{eqnarray}
&& u' = \frac{m'_1}{m+m_1} \left( \frac{m m_1}{m' m'_1} \right)^{1/2} \big[ v-v_1 - 2 (v-v_1) \cdot \Omega  \, \Omega \big], \label{express_|u'|} \\
&& u'_1 = - \frac{m'}{m+m_1} \left( \frac{m m_1}{m' m'_1} \right)^{1/2} \big[ v-v_1 - 2 (v-v_1) \cdot \Omega  \, \Omega \big] \, \Omega. \label{express_|u1'|} 
\end{eqnarray}
This results in the following collision law: 
\begin{eqnarray}
&& v' = \frac{m v + m_1 v_1}{m+m_1} + \frac{(m m_1)^{1/2}}{m+m_1} \left( \frac{m'_1}{m'}\right)^{1/2} \big[ v-v_1 - 2 (v-v_1) \cdot \Omega  \, \Omega \big], \label{express_v'} \\
&& v'_1 = \frac{m v + m_1 v_1}{m+m_1} - \frac{(m m_1)^{1/2}}{m+m_1} \left( \frac{m'}{m'_1}\right)^{1/2} \big[ v-v_1 - 2 (v-v_1) \cdot \Omega  \, \Omega \big], \label{express_v1'}  \\
&& m'_1 = m+m_1-m', \label{express_m1'}
\end{eqnarray}
where $m'$ takes any integer value $\{1, 2, \ldots, m+m_1-1 \}$ and $\Omega$ in ${\mathbb S}^{n-1}$ with $\Omega \cdot (v-v_1) \leq 0$. 

Eq. \eqref{eq:v'-v1'} can be inverted into 
\begin{equation} 
(m m_1)^{1/2} (v-v_1) = (m' m'_1)^{1/2} \big[ v'-v'_1 - 2 (v'-v'_1) \cdot \Omega  \, \Omega \big], 
\label{eq:v-v1}
\end{equation}
which leads to 
\begin{eqnarray}
&& v = \frac{m' v' + m'_1 v'_1}{m'+m'_1} + \frac{(m' m'_1)^{1/2}}{m'+m'_1} \left( \frac{m_1}{m}\right)^{1/2} \big[ v'-v'_1 - 2 (v'-v'_1) \cdot \Omega'  \, \Omega' \big], \label{express_v} \\
&& v_1 = \frac{m' v' + m'_1 v'_1}{m'+m'_1} - \frac{(m' m'_1)^{1/2}}{m'+m'_1} \left( \frac{m}{m_1}\right)^{1/2} \big[ v'-v'_1 - 2 (v'-v'_1) \cdot \Omega'  \, \Omega' \big], \label{express_v1}  \\
&& m_1 = m'+m'_1-m, \label{express_m1}
\end{eqnarray}
with $\Omega' = -\Omega$ and we note that $\Omega' \in {\mathbb S}^{n-1}$ satisfies $\Omega' \cdot (v'-v'_1) \leq 0$.

\subsection{The Boltzmann operator with mass exchange (BME operator)}
\label{subsec_boltzmann}

Let $f_m(x,v,t)$ be the distribution function of particles of mass $m$, and $f = (f_m)_{m=1, \ldots }$. We define a Boltmann collision operator for collisions with mass exchange (in short BME operator) in weak sense. Let $\varphi_m(v)$ be a sequence of test functions of the variable $v$. Then, we write
\begin{eqnarray}
&&\hspace{-1cm}
\frac{d}{dt} \Big( \sum_{m=1}^\infty \int_{{\mathbb R}^n} f_m \, \varphi_m \, dv \Big) =  \sum_{m=1}^\infty \int_{{\mathbb R}^n} Q_m(f) (v) \varphi_m \, dv  
= \frac{1}{2} \sum_{m,m_1=1}^\infty \sum_{m'=1}^{m+m_1-1} A_{m,m_1;m'} \nonumber \\
&&\hspace{0cm}
\int_{(v,v_1,\Omega) \in {\mathbb R}^{2n} \times {\mathbb S}^{n-1} \, | \, (v-v_1) \cdot \Omega \leq 0} {\mathbf B} \Big( \frac{m \, m_1}{m+m_1} |v-v_1|^2, \Omega \cdot \frac{v-v_1}{|v-v_1|} \Big) \, \nonumber \\
&&\hspace{5cm}
 \, \big( \varphi_{m'} + \varphi_{m'_1} - \varphi_m - \varphi_{m_1} \big) \, f_m \, f_{m_1} \, dv \, dv_1 \, d\Omega, 
\label{eq:Boltz_weak} 
\end{eqnarray}
where $\varphi_m = \varphi_m(v)$, $\varphi_{m_1} = \varphi_{m_1}(v_1)$, $\varphi_{m'} = \varphi_{m'}(v')$, $\varphi_{m'_1} = \varphi_{m'_1}(v'_1)$, and similarly for $f_m$, $f_{m_1}$, $f_{m'}$, $f_{m'_1}$ and $v'$, $v'_1$, $m'_1$ related to $v$, $v_1$, $m$, $m_1$, $m'$ and $\Omega$ through \eqref{express_v'}-\eqref{express_m1'}. Here, $A_{m,m_1;m'}$ is the rate of collisions between particles of masses $m$ and $m_1$ giving rise to particles of masses $m'$ and $m+m_1-m'$, while ${\mathbf B}$ is the rate of collisions of particles $(m,v)$ with particles $(m_1,v_1)$ generating a deflection angle parametrized by $\Omega$. For hard-sphere collisions, the number of collisions undergone by a test particle $(m,v)$ against field particles $(m_1,v_1)$ during a time interval $dt$ is equal to the number of such field particles in a collision cylinder whose radius $r_c$ is the sum of the radii of the two particles, and of height equal to $|v-v_1| dt$. The radius of a particle of mass $m$ scales like $m^{1/n}$, so that we can suppose that the surface of the section of this cylinder is $\sigma = C_n (m^{1/n} + m_1^{1/n})^{n-1}$, where $C_n$ is the volume of the unit ball in an $(n-1)$-dimensional euclidean space. We note that we can aggregate the factor $C_n (m^{1/n} + m_1^{1/n})^{n-1}$ with the constant $A_{m,m_1;m'}$ and assume that $\sigma = 1$. For more general collisions, the number $\sigma$ is replaced by the so-called differential scattering cross section, which is classically a  function of the kinetic energy of the reduced particle $E_{\textrm{red}} = \frac{m \, m_1}{m+m_1}|v-v_1|^2$,  and of $\Omega \cdot \frac{v-v_1}{|v-v_1|}$ which is the cosine of $\pi/2$ plus half the scattering angle. This is exact up to the same multiplicative factor as in the hard-sphere case, which again can be aggregated with the constant $A_{m,m_1;m'}$. Likewise, we can write $|v-v_1| = (\frac{m+m_1}{mm_1})^{1/2} E_{\textrm{red}}^{1/2} $ and the prefactor $(\frac{m+m_1}{mm_1})^{1/2}$ can be aggregated with $A_{m,m_1;m'}$. In the end, this shows that we can assume that the collision rate ${\mathbf B}$ is a function of $E_{\textrm{red}}$ and $\Omega \cdot \frac{v-v_1}{|v-v_1|}$, without any other dependence on the masses, which is what appears in \eqref{eq:Boltz_weak}. Of course, all this assumes factorization between rates for mass exchanges $A_{m,m_1;m'}$ and rates for velocity changes ${\mathbf B}$ ; in  total generality we should have postulated a rate of the form ${\mathbf B}_{m,m_1;m'} ( \frac{m \, m_1}{m+m_1} |v-v_1|^2, \Omega \cdot \frac{v-v_1}{|v-v_1|})$. This would lead to further complications that we wish to avoid in the present work.  

Since the partition of $m+m_1$ into $m$ and $m_1$ is the same as that obtained by exchanging $m$ and $m_1$, and similarly, the partition of $m+m_1$ into $m'$ and $m'_1$ is the same as that obtained by exchanging $m'$ and $m'_1$, we have 
\begin{equation}
A_{m,m_1;m'} = A_{m_1,m;m'} = A_{m,m_1;m+m_1-m'}, \, \,  \forall m, \, m_1 \geq 1, \, \,  \forall m' \in \{1, \ldots, m+m_1-1 \}. 
\label{eq:exchangability}
\end{equation} 
We also note that the parameter values $(m,v,m_1,v_1,\Omega,m')$ and $(m_1,v_1,m,v,-\Omega,m+m_1-m')$ correspond to the same physical collision. Thus, in the sums and integrals involved in \eqref{eq:Boltz_weak}, each physical collision is counted twice, which justifies the factor $1/2$ in front of the whole expression. 

We now give the strong form of the BME operator given in weak form by \eqref{eq:Boltz_weak}: 

\begin{lemma}[BME operator in strong form]
We have 
$$ \frac{d}{dt} f_m(v) = Q_m(f)(v), \quad \forall m \in \{1, 2, \ldots, \}, \quad \forall v \in {\mathbb R}^n, $$
with
\begin{eqnarray}
&&\hspace{-1cm}
Q_m(f)(v) = \int_{(v_1,\Omega) \in {\mathbb R}^n \times {\mathbb S}^{n-1} \, | \, (v-v_1) \cdot \Omega \leq 0} {\mathbf B} \Big( \frac{m \, m_1}{m+m_1} |v-v_1|^2, \Omega \cdot \frac{v-v_1}{|v-v_1|} \Big) \, \nonumber \\
&&\hspace{-1cm}
 \, \bigg[ \sum_{\{ (m',m'_1) \, | \, m'+m'_1 \geq m+1 \}} A_{m',m'_1;m} \, f_{m'} \, f_{m'_1} \Big( \frac{m \, m_1}{m' m'_1} \Big)^{\frac{n}{2}}  \nonumber \\
&&\hspace{6cm}
- \sum_{m_1=1}^\infty \sum_{m'=1}^{m+m_1-1} A_{m,m_1;m'} \, f_m \, f_{m_1} \bigg] \, dv_1 \, d\Omega, \label{eq:Boltz_strong}
\end{eqnarray}
\label{lem_Boltz_strong}
\end{lemma}

\noindent
\textbf{Proof.} The right-hand side of \eqref{eq:Boltz_weak}  is the sum of four terms labeled \textcircled{1} to \textcircled{4} in the same order as the terms associated with $\varphi_{m'}$, $\varphi_{m'_1}$, $\varphi_m$, $\varphi_{m_1}$. 
We immediately get 
\begin{eqnarray} 
&&\hspace{-1cm}
\textrm{\textcircled{3}} = - \frac{1}{2} \sum_{m=1}^\infty \int_{v \in {\mathbb R}^n} \bigg( \sum_{m_1=1}^\infty \sum_{m'=1}^{m+m_1-1} A_{m,m_1;m'} \, \nonumber\\
&&\hspace{-1.2cm}
\, \int_{(v_1,\Omega) \in {\mathbb R}^n \times {\mathbb S}^{n-1} \, | \, (v-v_1) \cdot \Omega \leq 0} {\mathbf B} \Big( \frac{m \, m_1}{m+m_1} |v-v_1|^2, \Omega \cdot \frac{v-v_1}{|v-v_1|} \Big) f_m \, f_{m_1} \, dv_1 \, d \Omega \bigg) \varphi_m \, dv. \label{eq:express_3}
\end{eqnarray}
Now, for \textcircled{4}, we just exchange $(m,v)$ and $(m_1,v_1)$ on the one hand, and change the sign of~$\Omega$ on the other hand. Since $m$, $m_1$, $v$, $v_1$ are dummy variables, the result is unchanged and we readily see that 
\begin{equation}
\textrm{\textcircled{4}} = \textrm{\textcircled{3}}. \label{eq:express_4}
\end{equation}

Now, we turn ourselves to \textcircled{1}. Changing the order of the sums in $m$, $m_1$, $m'$, we get 
\begin{eqnarray} 
&&\hspace{-1cm}
\textrm{\textcircled{1}} = \frac{1}{2} \sum_{m'=1}^\infty  \, \, \sum_{\{ (m,m_1) \, | \, m+m_1 \geq m'+1 \}} A_{m,m_1;m'} \, \nonumber \\
&&\hspace{-1.5cm}
\, \int_{(v,v_1,\Omega) \in {\mathbb R}^{2n} \times {\mathbb S}^{n-1} \, | \, (v-v_1) \cdot \Omega \leq 0} {\mathbf B} \Big( \frac{m \, m_1}{m+m_1} |v-v_1|^2, \Omega \cdot \frac{v-v_1}{|v-v_1|} \Big) \varphi_{m'} \, f_m \, f_{m_1} \, dv \, dv_1 \, d \Omega . \label{eq:express_1_inter}
\end{eqnarray}
We now make the change of variables $(v,v_1) \to (v',v'_1)$. We need to compute the Jacobian $J = |\textrm{det} (\frac{\partial (v,v_1)}{\partial (v',v'_1)})| = |\textrm{det} (\frac{\partial (v',v'_1)}{\partial (v,v_1)})|^{-1}$. From \eqref{express_v'}, \eqref{express_v1'}, we note that the map $(v,v_1) \mapsto (v',v'_1)$ for given $\Omega$ is linear and can be written
\begin{eqnarray*}
v' &=& H v - K (v \cdot \Omega) \Omega + L v_1 + K (v_1 \cdot \Omega) \Omega, \\
v'_1 &=& H' v + K' (v \cdot \Omega) \Omega + L' v_1 - K' (v_1 \cdot \Omega) \Omega,
\end{eqnarray*}
with 
\begin{eqnarray*}
&&\hspace{-1cm}
H= \alpha + \sqrt{\alpha \alpha_1} \, \gamma, \quad K = 2 \sqrt{\alpha \alpha_1} \, \gamma, \quad L= \alpha_1 - \sqrt{\alpha \alpha_1} \, \gamma, \\
&&\hspace{-1cm}
H'= \alpha - \frac{\sqrt{\alpha \alpha_1}}{\gamma}, \quad K' = 2 \frac{\sqrt{\alpha \alpha_1}}{\gamma}, \quad L'= \alpha_1 + \frac{\sqrt{\alpha \alpha_1}}{\gamma}, 
\end{eqnarray*}
and 
$$ \alpha = \frac{m}{m+m_1}, \quad \alpha_1 = \frac{m_1}{m+m_1}, \quad \gamma = \sqrt{\frac{m'_1}{m'}}. $$
Let $(e_1, \ldots, e_n)$ be basis of ${\mathbb R}^n$ such that $e_n = \Omega$. Then, in this basis, the linear map $(v,v_1) \mapsto (v',v'_1)$ has matrix ${\mathcal X}$ given blockwise by: 
$$ {\mathcal X} = \left( \begin{array}{cc} {\mathcal A} & {\mathcal B} \\ {\mathcal C} & {\mathcal D} \end{array} \right), $$
with 
\begin{eqnarray*}
{\mathcal A} &=& \textrm{diag}(H,\ldots, H, H-K),  \qquad {\mathcal B} = \textrm{diag}(L,\ldots, L, L+K), \\
{\mathcal C} &=& \textrm{diag}(H',\ldots, H', H'+K'), \qquad {\mathcal D} = \textrm{diag}(L',\ldots, L', L'-K'), \\
\end{eqnarray*}
where $\textrm{diag}(a_1, \ldots, a_n)$ denotes the diagonal matrix with diagonal entries $a_1, \ldots, a_n$. Since ${\mathcal C}$  and ${\mathcal D}$ commutent, we have 
\begin{eqnarray*} 
\textrm{det}{\mathcal X} &=& \textrm{det} \big({\mathcal A}{\mathcal D} - {\mathcal B} {\mathcal C} \big) \\
&=& (H L'-L H')^{n-1} \big( (H-K)(L'-K')-(H' + K')(L+K) \big). 
\end{eqnarray*}
Remarking that $\alpha + \alpha_1 = 1$ and using \eqref{eq:cons_mass}, we compute
\begin{eqnarray*}
H L'-L H' &=& - \big( (H-K)(L'-K')-(H' + K')(L+K) \big) \\
&=& \sqrt{\alpha \alpha_1} \Big( \gamma + \frac{1}{\gamma} \Big) = \sqrt{\frac{m \, m_1}{m' m'_1}}, 
\end{eqnarray*}
which leads to 
$$ J = |\textrm{det} {\mathcal X}|^{-1} = \Big( \frac{m' m'_1}{m \, m_1} \Big)^{\frac{n}{2}}. $$
Thus, owing to \eqref{eq:cons_mass} and \eqref{eq:reduced_ener}, the change of variables $(v,v_1) \to (v',v'_1)$ and $\Omega \to \Omega' = - \Omega$ in \eqref{eq:express_1_inter} leads to
\begin{eqnarray*} 
&&\hspace{-1cm}
\textrm{\textcircled{1}} = \frac{1}{2} \sum_{m'=1}^\infty  \, \, \sum_{\{ (m,m_1) \, | \, m+m_1 \geq m'+1 \}} A_{m,m_1;m'} \,  \\
&&\hspace{-1.5cm}
\, \int_{(v',v'_1,\Omega') \in {\mathbb R}^{2n} \times {\mathbb S}^{n-1} \, | \, (v'-v'_1) \cdot \Omega' \leq 0} {\mathbf B} \Big( \frac{m' \, m'_1}{m'+m'_1} |v'-v'_1|^2, \Omega' \cdot \frac{v'-v'_1}{|v'-v'_1|} \Big) \, \\
&&\hspace{7cm}
\, \varphi_{m'} \, f_m \, f_{m_1} \, \Big( \frac{m' m'_1}{m \, m_1} \Big)^{\frac{n}{2}} \, dv' \, dv'_1 \, d \Omega' . 
\end{eqnarray*}
Now, we just rename $(m,v)$ into $(m',v')$ and vice-versa, and $(m_1,v_1)$ into $(m'_1, v'_1)$ and vice-versa and rename $\Omega'$ into $\Omega$. This gives 
\begin{eqnarray} 
&&\hspace{-1cm}
\textrm{\textcircled{1}} = \frac{1}{2}  \sum_{m=1}^\infty  \int_{v \in {\mathbb R}^{n}}  \bigg( \sum_{\{ (m',m'_1) \, | \, m'+m'_1 \geq m+1 \}} A_{m',m'_1;m} \nonumber  \\
&&\hspace{-1.5cm}
\, \int_{(v_1,\Omega) \in {\mathbb R}^{n} \times {\mathbb S}^{n-1} \, | \, (v-v_1) \cdot \Omega \leq 0} {\mathbf B} \Big( \frac{m \, m_1}{m+m_1} |v-v_1|^2, \Omega \cdot \frac{v-v_1}{|v-v_1|} \Big) \nonumber \\
&&\hspace{6.5cm}
\, f_{m'} \, f_{m'_1} \, \Big( \frac{m m_1}{m' \, m'_1} \Big)^{\frac{n}{2}} \, dv_1 \, d \Omega \bigg) \, \varphi_m \, dv . \label{eq:express_1}
\end{eqnarray}

Finally, using the same algebra as for \textcircled{1}, we get 
\begin{eqnarray*} 
&&\hspace{-1cm}
\textrm{\textcircled{2}} = \frac{1}{2} \sum_{m=1}^\infty  \, \, \sum_{\{ (m',m'_1) \, | \, m'+m'_1 \geq m+1 \}} A_{m',m'_1;m} \,  \\
&&\hspace{-1.5cm}
\, \int_{(v,v_1,\Omega) \in {\mathbb R}^{2n} \times {\mathbb S}^{n-1} \, | \, (v-v_1) \cdot \Omega \leq 0} {\mathbf B} \Big( \frac{m \, m_1}{m+m_1} |v-v_1|^2, \Omega \cdot \frac{v-v_1}{|v-v_1|} \Big) \, \\
&&\hspace{7cm}
\, \varphi_{m_1} \, f_{m'} \, f_{m'_1} \, \Big( \frac{m m_1}{m' \, m'_1} \Big)^{\frac{n}{2}} \, dv \, dv_1 \, d \Omega , 
\end{eqnarray*}
Now, as for \textcircled{4}, we exchange $(m,v)$ and $(m_1,v_1)$ on the one hand and change the sign of~$\Omega$ on the other hand and get
\begin{equation}
\textrm{\textcircled{2}} = \textrm{\textcircled{1}}. \label{eq:express_2}
\end{equation}

Adding \eqref{eq:express_3}, \eqref{eq:express_4}, \eqref{eq:express_1} and \eqref{eq:express_2},  and owing to the fact that the resulting formula is valid for any sequence of test functions $(\varphi_m)_{m \geq 1}$, we are led to \eqref{eq:Boltz_strong}. \endproof

\begin{lemma}[Equivalent weak form of the BME operator]
Eq. \eqref{eq:Boltz_weak} is equivalent to the following:
\begin{eqnarray}
&&\hspace{-1cm}
\sum_{m=1}^\infty \int_{{\mathbb R}^n} Q_m(f) (v) \varphi_m \, dv  = - \frac{1}{4} \sum_{m,m_1=1}^\infty \sum_{m'=1}^{m+m_1-1} (m \, m_1)^{\frac{n}{2}}  \nonumber \\
&&\hspace{-0.5cm}
\int_{(v,v_1,\Omega) \in {\mathbb R}^{2n} \times {\mathbb S}^{n-1} \, | \, (v-v_1) \cdot \Omega \leq 0} {\mathbf B} \Big( \frac{m \, m_1}{m+m_1} |v-v_1|^2, \Omega \cdot \frac{v-v_1}{|v-v_1|} \Big) \, \nonumber \\
&&\hspace{0cm}
\big( \varphi_{m'} + \varphi_{m'_1} - \varphi_m - \varphi_{m_1} \big) \, \bigg[  \frac{A_{m',m'_1;m}}{(m' m'_1)^{\frac{n}{2}}} \, f_{m'} \, f_{m'_1} - \frac{A_{m,m_1;m'}}{(m \,  m_1)^{\frac{n}{2}}} f_m \, f_{m_1} \bigg] \, dv \, dv_1 \, d\Omega.
\label{eq:Boltz_weak_2} 
\end{eqnarray}
\label{lem:Boltz_weak_2}
\end{lemma}

\noindent
\textbf{Proof.} The proof is similar to that of the previous lemma. We start with \eqref{eq:Boltz_weak} and make the change of variables $(m, v , m_1, v_1) \to (m', v', m'_1, v'_1)$ and $\Omega \to \Omega' = - \Omega$. We get 
\begin{eqnarray*}
&&\hspace{-1cm}
\sum_{m=1}^\infty \int_{{\mathbb R}^n} Q_m(f) (v) \varphi_m \, dv  
= \frac{1}{2} \sum_{m,m_1=1}^\infty \sum_{m'=1}^{m+m_1-1} A_{m,m_1;m'}  \\
&&\hspace{0cm}
\int_{(v',v'_1,\Omega') \in {\mathbb R}^{2n} \times {\mathbb S}^{n-1} \, | \, (v'-v'_1) \cdot \Omega' \leq 0} {\mathbf B} \Big( \frac{m' \, m'_1}{m'+m'_1} |v'-v'_1|^2, \Omega' \cdot \frac{v'-v'_1}{|v'-v'_1|} \Big) \\
&&\hspace{3cm}
 \, \big( \varphi_{m'} + \varphi_{m'_1} - \varphi_m - \varphi_{m_1} \big) \, f_m \, f_{m_1} \, \, \Big( \frac{m' m'_1}{m \, m_1} \Big)^{\frac{n}{2}} \, dv' \, dv'_1 \, d\Omega'. 
\end{eqnarray*}
Now, we rename $(m,v)$ into $(m',v')$ and vice-versa, and $(m_1,v_1)$ into $(m'_1, v'_1)$ and vice-versa and rename $\Omega'$ into $\Omega$. This gives 
\begin{eqnarray*}
&&\hspace{-1cm}
\sum_{m=1}^\infty \int_{{\mathbb R}^n} Q_m(f) (v) \varphi_m \, dv  
= - \frac{1}{2} \sum_{m',m'_1=1}^\infty \sum_{m=1}^{m'+m'_1-1} A_{m',m'_1;m}  \\
&&\hspace{0cm}
\int_{(v,v_1,\Omega) \in {\mathbb R}^{2n} \times {\mathbb S}^{n-1} \, | \, (v-v_1) \cdot \Omega \leq 0} {\mathbf B} \Big( \frac{m \, m_1}{m+m_1} |v-v_1|^2, \Omega \cdot \frac{v-v_1}{|v-v_1|} \Big) \\
&&\hspace{3cm}
 \, \big( \varphi_{m'} + \varphi_{m'_1} - \varphi_m - \varphi_{m_1} \big) \, f_{m'} \, f_{m'_1} \, \, \Big( \frac{m m_1}{m' \, m'_1} \Big)^{\frac{n}{2}} \, dv \, dv_1 \, d\Omega.
\end{eqnarray*}
Now, we note that 
$$ \sum_{m',m'_1=1}^\infty \sum_{m=1}^{m'+m'_1-1} = \sum_{m,m_1=1}^\infty \sum_{m'=1}^{m+m_1-1}. $$
Thus, we get 
\begin{eqnarray}
&&\hspace{-1cm}
\sum_{m=1}^\infty \int_{{\mathbb R}^n} Q_m(f) (v) \varphi_m \, dv  
= - \frac{1}{2} \sum_{m,m_1=1}^\infty \sum_{m'=1}^{m+m_1-1} A_{m',m'_1;m} \nonumber \\
&&\hspace{0cm}
\int_{(v,v_1,\Omega) \in {\mathbb R}^{2n} \times {\mathbb S}^{n-1} \, | \, (v-v_1) \cdot \Omega \leq 0} {\mathbf B} \Big( \frac{m \, m_1}{m+m_1} |v-v_1|^2, \Omega \cdot \frac{v-v_1}{|v-v_1|} \Big) \nonumber\\
&&\hspace{3cm}
 \, \big( \varphi_{m'} + \varphi_{m'_1} - \varphi_m - \varphi_{m_1} \big) \, f_{m'} \, f_{m'_1} \, \, \Big( \frac{m m_1}{m' \, m'_1} \Big)^{\frac{n}{2}} \, dv \, dv_1 \, d\Omega.
\label{eq:Boltz_weak_inter}
\end{eqnarray} 
Then, adding \eqref{eq:Boltz_weak} and \eqref{eq:Boltz_weak_inter} leads to \eqref{eq:Boltz_weak_2}, which ends the proof. \endproof

\subsection{Conservations, entropy dissipation and equilibria}
\label{subsec_conservations}

\begin{proposition}[Conservations]
Let $\varphi = (\varphi_m)_{m \geq 1}$ be a sequence of smooth functions with sufficient decay at infinity in $m$ and $v$ such that 
\begin{equation}
\varphi_{m'}(v') + \varphi_{m'_1}(v'_1) - \varphi_m(v) - \varphi_{m_1}(v_1) = 0, 
\label{eq:identity_coll_invar}
\end{equation}
for all $(m',v')$, $(m'_1,v'_1)$, $(m,v)$, $(m_1,v_1)$ satisfying \eqref{eq:cons_mass}, \eqref{eq:cons_momentum}, \eqref{eq:cons_energy}. Then, we have 
\begin{equation}
\sum_{m=1}^\infty \int_{{\mathbb R}^n} Q_m(f) \varphi_m \, dv = 0, 
\label{eq:conservations}
\end{equation}
for any sequence $f = (\varphi_m)_{m \geq 1}$ of smooth functions decaying fast enough in $v$ and $m$.  
\label{prop:conservations}
\end{proposition}

\noindent
\textbf{Proof.} We apply \eqref{eq:Boltz_weak_2} with $\varphi$ satisfying \eqref{eq:identity_coll_invar} and immediately get \eqref{eq:conservations}. 
\endproof

\begin{lemma}[Collisional invariants]
Any sequence $\varphi = (\varphi_m)_{m \geq 1}$ of smooth functions ${\mathbb R}^n \to {\mathbb R}$ satisfying \eqref{eq:identity_coll_invar} belongs to the space 
$$ {\mathcal C} = \textrm{span} \{ \mathbf{1}, \mathbf{m}, \mathbf{mv_1}, \ldots, \mathbf{mv_n}, \mathbf{m|v|^2} \}, $$
where 
\begin{eqnarray*}
\mathbf{1} &=& (\varphi_m)_{m \geq 1} \quad \textrm{with} \quad \varphi_m(v) = 1, \\
\mathbf{m} &=& (\varphi_m)_{m \geq 1} \quad \textrm{with} \quad \varphi_m(v) = m, \\
\mathbf{m v_k} &=& (\varphi_m)_{m \geq 1} \quad \textrm{with} \quad \varphi_m(v) = m v_k, \quad k \in \{1, \ldots, n \}, \\
\mathbf{m |v|^2} &=& (\varphi_m)_{m \geq 1} \quad \textrm{with} \quad \varphi_m(v) = m |v|^2.
\end{eqnarray*} 
\label{lem:invar_collision}
\end{lemma}

\noindent
\textbf{Proof.}  
The proof is divided into four steps.

\noindent \textbf{Step 1.} For fixed $m=m_1=m'=m'_1$, \eqref{eq:identity_coll_invar} becomes
\begin{equation*}
\varphi_m(v) + \varphi_{m}(v_1) = \varphi_m(v') + \varphi_{m}(v'_1), 
\end{equation*}
for all $v, v_1, v', v'_1$ satisfying $v+v_1=v'+v'_1$ and $|v|^2+|v_1|^2 =|v'|^2+|v'_1|^2$. Thanks to a classical result \cite[Prop 28.2]{tartar2008hyperbolic}, there exist constants $C_m$, $A_m$ and a constant vector $B_m$ depending on $m$ such that
\begin{equation}
\varphi_m(v) = C_m  |v|^2 + D_m \cdot  v + A_m.
\label{eq:inv_phi}
\end{equation}

\noindent \textbf{Step 2.} We demonstrate that $A_m = m B + A$ for some constants $A$ and $B$. Indeed, by setting  $v=v_1=v'=v'_1=0$, for $m+m_1=m'+m'_1$ and using equation \eqref{eq:inv_phi}, \eqref{eq:identity_coll_invar} becomes
\begin{equation}
 A_{m} + A_{m_1}= A_{m'} + A_{m'_1}.
 \label{eq:inv_C}
\end{equation}
For $m\ge 2$, since $m + 1= (m-1) + 2$  one gets  $A_m = A_{m-1} + (A_2 -A_1)$. 
Recursively, for $m\ge 3$, we get
\begin{equation}
A_m = A_2 + (m-2)(A_2-A_1)=m(A_2-A_1) + 2 A_1 - A_2. 
\end{equation}
This equation is also valid for $m=1$ and $2$. Thus, we can set $B=A_2-A_1$ and $A=2 A_1 - A_2$.

\noindent \textbf{Step 3.} We show that $C_m$ and $D_m$ are affine functions of $m$. For $m+m_1=m'+m'_1$, and $v=v_1=v'=v'_1$,  the conditions \eqref{eq:cons_momentum} and \eqref{eq:cons_energy} are met.  Using equations \eqref{eq:inv_phi} and \eqref{eq:inv_C},  then \eqref{eq:identity_coll_invar} reduces to 
\begin{equation}
C_{m}  |v|^2 + D_{m}\cdot v + C_{m_1} |v|^2 + D_{m_1}\cdot v= 
C_{m'} |v|^2 + D_{m'}\cdot v + C_{m'_1} \,  |v|^2 + D_{m'_1}\cdot v. 
\label{eq:inv_AB}
\end{equation}
Now, $2v=2v_1=2v'=2v'_1$ still satisfy \eqref{eq:cons_momentum} and \eqref{eq:cons_energy}. Thus, we have
\begin{equation}
4C_{m}  |v|^2 + 2D_{m}\cdot v + 4C_{m_1} |v|^2 + 2D_{m_1}\cdot v= 
4C_{m'} |v|^2 + 2D_{m'}\cdot v +4 C_{m'_1} \,  |v|^2 + 2D_{m'_1}\cdot v. 
\label{eq:inv_AB1}
\end{equation}
Equations \eqref{eq:inv_AB} and \eqref{eq:inv_AB1} yield
\begin{align*}
C_m+  C_{m_1}=C_{m'}  +  C_{m'_1},  \\
D_m+  D_{m_1}=D_{m'}  +  D_{m'_1}. 
\end{align*}
Using the same argument as in Step 2, we find  
\begin{equation}
C_m= m C + C_0, \quad D_m= m D + D_0. 
\label{eq:inv_AB2}
\end{equation}
for some constants $C$, $C_0$ and some constant vectors $D$, $D_0$.

\noindent \textbf{Step 4.} We now prove $C_0=0$ and $D_0=0$. Using equations \eqref{eq:inv_phi}, \eqref{eq:inv_C} and \eqref{eq:inv_AB2} together with \eqref{eq:cons_momentum} and \eqref{eq:cons_energy}, Eq.  \eqref{eq:identity_coll_invar} is reduced to  
\begin{align*}
C_0 (|v|^2+ |v_1|^2) +  D_0\cdot(v+ v_1) = C_0 (|v'|^2+ |v'_1|^2) +  D_0\cdot(v'+ v'_1). 
\end{align*}
Using again the factor 2 trick as in Step 3, we have
\begin{align}
 C_0 (|v|^2+ |v_1|^2) =C_0 (|v'|^2+ |v'_1|^2), \label{eq:inv_step4_A0}\\
 D_0\cdot(v+ v_1) =  D_0\cdot(v'+ v'_1). \label{eq:inv_step4_B0}
\end{align}
We first show that $D_0 = 0$. By subtracting $2 v_{\textrm{CM}}$ (where $v_{\textrm{CM}}$ is given by \eqref{eq:COM_vel}) to each side of \eqref{eq:inv_step4_B0} and using 
\eqref{eq:express_u}-\eqref{eq:express_u1'}, we get
\begin{equation}
\frac{m_1-m}{m_1+m} \, \big(D_0 \cdot  (v-v_1) \big)= \frac{m'_1-m'}{m'_1+m'} \, \big( D_0 \cdot (v'-v'_1) \big). 
\label{eq:inv_step4_B0_2}
\end{equation}
Now, we assume that $D_0 \not = 0$. We choose $m=m_1$, $m' \not = m'_1$ (this requires $m >1$), $v$, $v_1$ such that $D_0 \cdot (v-v_1) \not = 0$ and $\Omega$ such that $\Omega \cdot (v-v_1) = 0$ (see collision rule \eqref{express_v'}, \eqref{express_v1'}). Then, according to the collision rule, 
$$ v'-v'_1 = \frac{1}{2} \Big[ \big( \frac{m'_1}{m'} \big)^{1/2} + \big( \frac{m'}{m'_1} \big)^{1/2} \Big] \, (v-v_1), \quad \textrm{ and so, } \quad \big( D_0 \cdot (v'-v'_1) \big) \not = 0. $$
Thus, the left-hand side of \eqref{eq:inv_step4_B0_2} is $0$ while the right-hand side is different from $0$ given the choices made. This yields a contradiction, and it results that $D_0=0$. Now, we show that $C_0 = 0$. Indeed, applying \eqref{eq:inv_step4_A0} with $(v,v_1,v',v'_1)$ replaced by $(w+v, w+v_1, w+v', w+v'_1)$ where $w$ is an arbitrary vector of ${\mathbb R}^n$ (we easily check that this quadruple of vectors still satisfies \eqref{eq:cons_momentum} and \eqref{eq:cons_energy}), we deduce that $C_0$ must satisfy
$$ C_0 (v+ v_1) =  C_0 (v'+ v'_1). $$
Taking an arbitrary unit vector $\mathbf{n}$, we deduce that 
$$ C_0 \mathbf{n} \cdot  (v+ v_1) =  C_0 \mathbf{n} \cdot (v'+ v'_1), $$
which is \eqref{eq:inv_step4_B0} with $D_0$ replaced by $C_0 \mathbf{n} $. The proof done for $D_0$ shows that $C_0 \mathbf{n}  = 0$ which in turn, shows that $C_0=0$. 

Combining all the above steps, we derive the following form for $\varphi_m(v)$:
$$
\varphi_m(v) =  A + Bm + C m  |v|^2 + D \cdot  m v .
$$
This completes the proof of the lemma.
\endproof

We now make the following

\begin{hypothesis}[Multiplicative exchange rates]~

\noindent
We assume that there exists a sequence $(\gamma_m)_{m \geq 1}$ with $\gamma_m \in  [0,\infty)$, $\forall m \geq 1$, such that 
\begin{equation} A_{m,m_1;m'} = \gamma_m \, \gamma_{m_1}, \quad \forall m, \, m_1 \geq 1, \quad \forall m' \in \{1, \ldots, m+m_1-1 \}. 
\label{eq:multiplicative}
\end{equation}
\label{hyp:multiplicative}
\end{hypothesis}

We remark that $A_{m,m_1;m'}$ given by \eqref{eq:multiplicative} satisfies \eqref{eq:exchangability}. 

\begin{proposition}[Entropy dissipation]~

\noindent
Assuming \eqref{eq:multiplicative}, we get 
\begin{equation}
\sum_{m=1}^\infty \int_{{\mathbb R}^n} Q_m(f) \log \Big(\frac{\gamma_m f_m}{m^\frac{n}{2}} \Big) \, dv \leq 0.
\label{eq:entropy_decay}
\end{equation}
\label{prop:entropy_decay}
\end{proposition}

\noindent
\textbf{Proof.} Taking 
\begin{equation}
\varphi_m = \log \Big(\frac{\gamma_m f_m}{m^\frac{n}{2}} \Big), 
\label{eq:phim=logfm}
\end{equation}
and using \eqref{eq:multiplicative}, Expression \eqref{eq:Boltz_weak_2}  becomes
\begin{eqnarray}
&&\hspace{-1cm}
\sum_{m=1}^\infty \int_{{\mathbb R}^n} Q_m(f) (v) \varphi_m \, dv  = - \frac{1}{4} \sum_{m,m_1=1}^\infty \sum_{m'=1}^{m+m_1-1} (m \, m_1)^{\frac{n}{2}} \nonumber \\
&&\hspace{1cm}
\int_{(v,v_1,\Omega) \in {\mathbb R}^{2n} \times {\mathbb S}^{n-1} \, | \, (v-v_1) \cdot \Omega \leq 0} {\mathbf B} \Big( \frac{m \, m_1}{m+m_1} |v-v_1|^2, \Omega \cdot \frac{v-v_1}{|v-v_1|} \Big) \nonumber \\
&&\hspace{3cm}
\bigg[ \log \Big( \frac{\gamma_{m'} f_{m'}}{(m')^{\frac{n}{2}}} \, \frac{\gamma_{m'_1} f_{m'_1}}{(m'_1)^{\frac{n}{2}}} \Big) - \log \Big( \frac{\gamma_m f_m}{m^{\frac{n}{2}}} \, \frac{\gamma_{m_1} f_{m_1}}{(m_1)^{\frac{n}{2}}} \Big) \bigg] \nonumber \\
&&\hspace{5cm}
 \bigg[  \frac{\gamma_{m'} f_{m'}}{(m')^{\frac{n}{2}}} \, \frac{\gamma_{m'_1} f_{m'_1}}{(m'_1)^{\frac{n}{2}}} - \frac{\gamma_m f_m}{m^{\frac{n}{2}}} \, \frac{\gamma_{m_1} f_{m_1}}{(m_1)^{\frac{n}{2}}} \bigg] \, dv \, dv_1 \, d\Omega, \label{eq:sum_int_Qm_phim}
\end{eqnarray}
and we obtain \eqref{eq:entropy_decay} owing to the fact that the logarithm is an increasing function. \endproof

We now make the following assumption about the growth of $\gamma_m$ with $m$: 

\begin{hypothesis}[Behaviour of $\gamma_m$ as $m \to \infty$]~

\noindent
Let ${\mathcal S}$ be the set
\begin{equation} {\mathcal S} = \Big\{ \beta \in {\mathbb R} \quad \Big| \quad \sum_{m \geq 1} \frac{m e^{\beta m}}{\gamma_m} < \infty \Big\}. 
\label{eq:def_S}
\end{equation}
We assume that ${\mathcal S} \not = \emptyset$. 
\label{hyp:gamma_m_to_infty}
\end{hypothesis}

\begin{remark}
As a consequence of Hypothesis \ref{hyp:gamma_m_to_infty}, there exists $\beta_0 \in {\mathbb R} \cup \{ + \infty \}$ such that ${\mathcal S}$ is of the form ${\mathcal S} = (- \infty, \beta_0)$ or ${\mathcal S} = (- \infty, \beta_0]$. 
\end{remark}

\begin{proposition}[Equilibria]~

\noindent
(i) Let $f$ having finite mass, momentum and energy, i.e. such that 
\begin{equation} 
\sum_{m \geq 1} \int_{{\mathbb R}^n} f_m(v) \, m (1 + |v|^2) \, dv < \infty. 
\label{eq:finite_moments}
\end{equation}
Under Hypotheses \ref{hyp:multiplicative} and \ref{hyp:gamma_m_to_infty}, we have $Q(f) = 0$ if and only if $\exists (\rho, \beta, u, \Theta)  \in [0,\infty) \times {\mathcal S} \times {\mathbb R}^n \times [0,\infty)$ such that 
\begin{equation}
f = \rho M_{u,\Theta, \beta} 
\label{eq:equilibria_0}
\end{equation}
with 
\begin{equation}
(M_{u,\Theta, \beta})_m(v) = \frac{1}{Z(\beta, \Theta)} \frac{m^{\frac{n}{2}} e^{\beta m}}{\gamma_m} \exp \Big( - \frac{m|v - u|^2}{2 \Theta} \Big), \quad \forall m \geq 1, \quad \forall v \in {\mathbb R}^n,  
\label{eq:equilibria}
\end{equation}
where 
\begin{equation}
Z(\beta, \Theta) = (2 \pi \Theta)^{\frac{n}{2}} \sum_{m \geq 1}  \frac{m e^{\beta m}}{\gamma_m}. 
\label{eq:normalization}
\end{equation}

\smallskip
\noindent
(ii) We have 
\begin{equation}
\sum_{m \geq 1} \int_{{\mathbb R}^n} \rho \big(M_{u,\Theta, \beta}\big)_m (v) \left( \begin{array}{c} 1 \\ m \\ mv \\ m |v-u|^2 \end{array} \right) \, dv = \left( \begin{array}{c} \rho \langle m^{-1} \rangle_\beta \\ \rho \\ \rho u \\  n \rho \Theta \langle m^{-1} \rangle_\beta \end{array} \right), 
\label{eq:moments_equi}
\end{equation} 
where for a sequence $a_m$ of real numbers, we denote by $\langle a_m \rangle_\beta$ the average
$$ \langle a_m \rangle_\beta = \frac{\displaystyle \sum_{m=1}^\infty \frac{m e^{\beta m}}{\gamma_m} a_m}{\displaystyle \sum_{m=1}^\infty \frac{m e^{\beta m}}{\gamma_m}}. $$
We note that $\langle a_m \rangle_\beta$ does not depend on $m$  and that it is a function of $\beta$. 
\label{prop:equilibria}
\end{proposition}

\textbf{Proof.} Suppose $Q(f)=0$. Then, 
\begin{equation}
\sum_{m=1}^\infty \int_{{\mathbb R}^n} Q_m(f) \log \Big(\frac{\gamma_m f_m}{m^\frac{n}{2}} \Big) \, dv = 0.
\label{eq:sum_int_Qm_logfm}
\end{equation}
We compute the left-hand side of \eqref{eq:sum_int_Qm_logfm} thanks to \eqref{eq:sum_int_Qm_phim}. Because the function inside the sum over $m$ and the integral over $v$ in \eqref{eq:sum_int_Qm_phim} is nonnegative, \eqref{eq:sum_int_Qm_logfm} implies that the function $\varphi_m$ given by \eqref{eq:phim=logfm} satisfies \eqref{eq:identity_coll_invar}. Hence, by Lemma \ref{lem:invar_collision}, there exist $A$, $B$, $C \in {\mathbb R}$ and $D \in {\mathbb R}^n$ such that 
\begin{equation} 
\log \Big(\frac{\gamma_m f_m}{m^\frac{n}{2}} \Big) = A + Bm + Cm|v|^2 + m D \cdot v, \quad \forall m \geq 1, \quad \forall v \in {\mathbb R}^n. 
\label{eq:maxwell_entropic_var}
\end{equation}
Integrability of $f_m$ with respect to $v \in {\mathbb R}^n$ requires $C <0$. We deduce that 
$$ f_m(v) = \alpha \, \frac{m^\frac{n}{2} e^{\beta m}}{\gamma_m} \, \exp \Big( - \frac{m |v-u|^2}{2 \Theta} \Big), $$
with 
\begin{equation} 
\alpha = e^A >0, \quad \beta = B - \frac{|D|^2}{4C}, \quad \Theta = - \frac{1}{2C} >0, \quad u = - \frac{D}{2C}. 
\label{eq:map_entro2cons}
\end{equation}
Then, using \eqref{eq:moments_maxwellian_1} from Appendix \ref{app_remark_form} with $p=0, \, 1$, we have 
$$ \sum_{m \geq 1} \int_{{\mathbb R}^n} f_m(v) \, m (1  + |v-u|^2) \, dv = \alpha \, \big( 2 \pi \Theta \big)^{\frac{n}{2}} \sum_{m \geq 1} \frac{e^{\beta m}}{\gamma_m} (m + n \Theta). $$
Thus, we have 
\begin{equation} 
\sum_{m \geq 1} \int_{{\mathbb R}^n} f_m(v) \, m (1  + |v-u|^2) \, dv < \infty \quad \Longleftrightarrow \quad \beta \in {\mathcal S} . 
\label{eq:equi_integrability}
\end{equation}
The statement at the left-hand side of the equivalence symbol in \eqref{eq:equi_integrability} is clearly equivalent to the statement \eqref{eq:finite_moments}. 
Now, using \eqref{eq:moments_maxwellian_1} with $p=0$ again together with \eqref{eq:normalization}, we get 
\begin{equation} 
\rho =: \sum_{m \geq 1} m \int_{{\mathbb R}^n} f_m(v) \, dv =  \alpha \, Z(\beta, \Theta), 
\label{eq:rel_rho_alpha}
\end{equation}
which leads to the expression \eqref{eq:equilibria} of $f_m$ as well as to the second line of the vector equation~\eqref{eq:moments_equi}. By antisymmetry, we immediately get 
$$ \sum_{m \geq 1} \int_{{\mathbb R}^n} f_m(v) \, m (v-u) \, dv = 0, $$
which directly leads to the third line of \eqref{eq:moments_equi}. Again, using \eqref{eq:moments_maxwellian_1} with $p=1$, we get 
$$ \sum_{m \geq 1} \int_{{\mathbb R}^n} f_m(v) \, m |v-u|^2 \, dv = n \Theta \, \frac{\rho}{Z(\beta, \Theta)} \, \big( 2 \pi \Theta \big)^{\frac{n}{2}} \sum_{m \geq 1} \frac{e^{\beta m}}{\gamma_m} = n \rho \Theta \langle m^{-1} \rangle_\beta.  $$
Finally, the first line of \eqref{eq:moments_equi} is obtained in the same way, using \eqref{eq:moments_maxwellian_1} with $p=0$. \endproof

\setcounter{equation}{0}
\section{Hydrodynamics of mass exchange processes}
\label{sec_Boltz_eq}

\subsection{The Euler system for mass exchange processes}
\label{subsec_EME}

The Boltzmann equation for mass exchange processes (or BME equation) is written 
\begin{equation}
\partial_t f_m^\varepsilon + v \cdot \nabla_x f_m^\varepsilon = \frac{1}{\varepsilon} Q_m(f^\varepsilon), \quad \forall m \geq 1, \quad \forall v \in {\mathbb R}^n, 
\label{eq:Boltz_eq}
\end{equation}
where $f^\varepsilon = (f_m^\varepsilon)_{m \geq 1}$ and where $Q_m$ is defined as in Section \ref{subsec_boltzmann}. Throughout this section, we assume that Hypotheses \ref{hyp:multiplicative} and \ref{hyp:gamma_m_to_infty} are fulfilled. Furthermore, we suppose that $f^\varepsilon|_{t=0} = f^\varepsilon_I$ is independent of $\varepsilon$. The goal of this section is to study the hydrodynamic limit $\varepsilon \to 0$. We have the following

\begin{theorem}[EME system in conservative form]
Suppose $f^\varepsilon$ exists on a time-interval $[0,T]$ independent of $\varepsilon$ and depends smoothly on $\varepsilon$. Then, as $\varepsilon \to 0$, $f^\varepsilon \to f = \rho M_{u,\Theta,\beta}$ where $(\rho,u,\Theta, \beta)$: ${\mathbb R}^n \times [0,T] \to [0,\infty) \times {\mathbb R}^n \times [0,\infty) \times {\mathcal S}$ satisfies the following system of PDE: 
\begin{equation}
\partial_t \left( \begin{array}{c} 
\rho \langle m^{-1} \rangle_\beta \\ \rho \\ \rho u \\ \rho |u|^2 + n \rho \Theta \langle m^{-1} \rangle_\beta 
\end{array} \right) + \nabla_x \cdot \left( \begin{array}{c} 
\rho u \langle m^{-1} \rangle_\beta \\ \rho u \\ \rho u \otimes u + \rho \Theta \langle m^{-1} \rangle_\beta \mathrm{I}_n \\ \rho |u|^2 u + (n+2) \rho \Theta \langle m^{-1} \rangle_\beta u \end{array} \right) = 0, 
\label{eq:hydro_limit}
\end{equation}
where $\mathrm{I}_n$ stands for the $n \times n$ identity matrix. This system will be referred to as the Euler system for mass exchange processes (EME). 
\label{th:hydro_limit}
\end{theorem}

\noindent
\textbf{Proof.} From \eqref{eq:Boltz_eq}, we have $Q(f^\varepsilon) = {\mathcal O}(\varepsilon)$. So, in the limit $\varepsilon \to 0$, we have $Q(f) = 0$, hence, $f = \rho M_{u, \Theta, \beta}$, where $(\rho,u,\Theta, \beta)$ may depend on $(x,t)$. Then, owing to Prop. \ref{prop:conservations}, we have 
$$
\partial_t \bigg( \sum_{m \geq 1} \int_{{\mathbb R}^n} f_m^\varepsilon 
\left( \begin{array}{c} 1 \\ m \\ mv \\ m|v|^2 \end{array} \right) \, dv \bigg) + \nabla_x \cdot \bigg( \sum_{m \geq 1} \int_{{\mathbb R}^n} f_m^\varepsilon 
\left( \begin{array}{c} 1 \\ m \\ mv \otimes \\ m|v|^2 \end{array} \right) \, v \, dv \bigg) = 0, 
$$
and, taking the limit $\varepsilon \to 0$, we get 
$$
\partial_t \bigg( \sum_{m \geq 1} \int_{{\mathbb R}^n} \rho (M_{u, \Theta, \beta})_m 
\left( \begin{array}{c} 1 \\ m \\ mv \\ m|v|^2 \end{array} \right) \, dv \bigg) + \nabla_x \cdot \bigg( \sum_{m \geq 1} \int_{{\mathbb R}^n} \rho (M_{u, \Theta, \beta})_m 
\left( \begin{array}{c} 1 \\ m \\ mv  \otimes \\ m|v|^2 \end{array} \right) \, v \, dv \bigg) = 0.  
$$
Thanks to \eqref{eq:moments_equi}, the first term gives the first term of \eqref{eq:hydro_limit}. For the second term, we have the following: 

\medskip
\noindent
\textbf{First line.} Since $(M_{u, \Theta, \beta})_m$ is even with respect to $v-u$, we have by antisymmetry with respect to $v-u$: 
$$\int_{{\mathbb R}^n} (M_{u, \Theta, \beta})_m(v) \, v \, dv = u \int_{{\mathbb R}^n} (M_{u, \Theta, \beta})_m(v) \, dv, $$ 
and 
$$ \sum_{m \geq 1} \int_{{\mathbb R}^n} \rho (M_{u, \Theta, \beta})_m(v) \, dv = \rho \langle m^{-1} \rangle_\beta, $$
by \eqref{eq:equilibria}, \eqref{eq:normalization} and \eqref{eq:moments_maxwellian_1} with $p=0$. 

\medskip
\noindent
\textbf{Second line.} The second line of the second term is identical with the third line of the first term, so, its value is $\rho u$. 

\medskip
\noindent
\textbf{Third line.} We have 
$$ v \otimes v = (v-u) \otimes (v-u) + u \otimes u + \textrm{ odd terms in } (v-u). $$
By antisymmetry, upon multiplication by $(M_{u, \Theta, \beta})_m(v)$ and integration with respect to $v$, only the even terms with respect to $v-u$ in the expression of $v \otimes v$ are  non-zero. With \eqref{eq:equilibria}, \eqref{eq:normalization} and \eqref{eq:moments_maxwellian_2} used with $p=0$, we have 
$$ \sum_{m \geq 1} \int_{{\mathbb R}^n} \rho (M_{u, \Theta, \beta})_m(v) \, m (v-u) \otimes (v-u) \, dv = \rho \Theta \langle m^{-1} \rangle_\beta \mathrm{I}_n, $$
while 
$$ \sum_{m \geq 1} \int_{{\mathbb R}^n} \rho (M_{u, \Theta, \beta})_m(v) \, m  dv = \rho, $$ by the second line of \eqref{eq:moments_equi}. So, the third line is $\rho (u \otimes u) + \rho \Theta \langle m^{-1} \rangle_\beta \mathrm{I}_n$. 

\medskip
\noindent
\textbf{Fourth line.} We write 
$$|v|^2 v = |v-u|^2 u + 2 \big( (v-u) \otimes (v-u) \big) u + |u|^2 u + \textrm{ odd terms in } (v-u), $$
and use again antisymmetry with respect to $v-u$ to eliminate the contribution of the odd terms. By the second and fourth lines of \eqref{eq:moments_equi}, we have 
$$ \sum_{m \geq 1} \int_{{\mathbb R}^n} \rho (M_{u, \Theta, \beta})_m(v) \, m |u|^2 u  dv = \rho |u|^2 u, $$
and 
$$ \sum_{m \geq 1} \int_{{\mathbb R}^n} \rho (M_{u, \Theta, \beta})_m(v) \, m |v-u|^2 u  dv = n \rho \Theta \langle m^{-1} \rangle u, $$
respectively. Finally, using \eqref{eq:moments_maxwellian_2} with $p=0$, we get
$$ 2 \sum_{m \geq 1} \int_{{\mathbb R}^n} \rho (M_{u, \Theta, \beta})_m(v) \, m \big( (v-u) \otimes (v-u) \big) u  dv = 2 \rho \Theta \langle m^{-1} \rangle u. $$
So, the fourth line equals $\rho |u|^2 u + (n+2) \rho \Theta \langle m^{-1} \rangle u$, which ends the proof. \endproof

Now, introducing $\tilde \Theta = \Theta \langle m^{-1} \rangle_\beta$, we can re-write the second to fourth lines of \eqref{eq:hydro_limit} into 
\begin{equation}
\partial_t \left( \begin{array}{c} 
 \rho \\ \rho u \\ \rho |u|^2 + n \rho \tilde \Theta  
\end{array} \right) + \nabla_x \cdot \left( \begin{array}{c} 
\rho u \\ \rho u \otimes u + \rho \tilde \Theta \mathrm{I}_n \\ \rho |u|^2 u + (n+2) \rho \tilde \Theta u \end{array} \right) = 0, 
\label{eq:hydro_limit_2}
\end{equation}
which is nothing but the standard gas dynamics equations for a perfect gas. Furthermore, combining the first and second line of \eqref{eq:hydro_limit}, the equation for $\beta$ reads 
$$
\partial_t \langle m^{-1} \rangle_\beta + u \cdot \nabla_x \langle m^{-1} \rangle_\beta = 0. 
$$
In fact, since the function $\beta \mapsto \langle m^{-1} \rangle_\beta$ is smooth, this equation can be turned into a direct equation for $\beta$: 
\begin{equation}
\partial_t \beta + u \cdot \nabla_x \beta = 0. 
\label{eq:beta_eq}
\end{equation}
Hence, the equations for the hydrodynamic quantities $(\rho, u, \tilde \Theta)$ are decoupled from the equation for the mass statistics $\beta$: once $u$ is known from the resolution of \eqref{eq:hydro_limit_2}, $\beta$ is obtained by solving \eqref{eq:beta_eq}. Of course, \eqref{eq:beta_eq} is ill-behaved at shocks and in such case, the use of the conservative equation resulting from the first line of \eqref{eq:hydro_limit} is preferable. The mass statistics though is crucial at initialization since it is needed to construct the initial value of the pseudo-temperature $\tilde \Theta$. Also, note that formulation \eqref{eq:hydro_limit_2}, \eqref{eq:beta_eq} tells us that the EME system is hyperbolic. The additional equation \eqref{eq:beta_eq} only increases the multiplicity of the eigenvalue $u$ by $1$. 

After some classical manipulations, we have the following

\begin{proposition}[EME system in nonconservative form]
For smooth solutions, the EME system \eqref{eq:hydro_limit} is equivalent to the following system 
\begin{eqnarray}
&&\hspace{-1cm}
\partial_t \beta + (u \cdot \nabla_x) \beta = 0, \label{eq:pop_NC} \\
&&\hspace{-1cm}
\partial_t \rho + (u \cdot \nabla_x) \rho + \rho (\nabla_x \cdot u) = 0, \label{eq:mass_NC} \\
&&\hspace{-1cm}
\partial_t u + (u \cdot \nabla_x) u + \frac{1}{\rho} \nabla_x \big( \rho \Theta \langle m^{-1} \rangle_\beta \big) = 0, \label{eq:mom_NC} \\
&&\hspace{-1cm}
\partial_t \Theta + (u \cdot \nabla_x) \Theta + \frac{2}{n} \Theta (\nabla_x \cdot u) = 0, \label{eq:temp_NC} 
\end{eqnarray}
which will be later referred to as the EME system in non-conservative form (while the original one \eqref{eq:hydro_limit} is the EME system in conservative form). 
\label{prop:EME_NC}
\end{proposition}

\subsection{The Navier-Stokes mass-exchange model}
\label{subsec_NS}

We now seek to compute the order $\varepsilon$ corrections to the EME model \eqref{eq:hydro_limit} using the Chapman-Enskog procedure. This procedure leads to diffusive terms. Compared with the classical Navier-Stokes equation, we expect additional diffusive corrections proportional to the gradients of the quantity $\beta$ involved in the equilibrium distribution $M_{u, \Theta, \beta}$. Since the computations can be quite tedious, we replace the BME operator by a BGK-type relaxation operator (below referred to as the BGKME operator) which has the same equilibria. So, for a given $f = (f_m(v))_{m \geq 1}$, the quadruple $(\rho_f, u_f, \beta_f, \Theta_f)$ is uniquely defined by the identity 
\begin{equation}
\left( \begin{array}{c} 
\rho_f \langle m^{-1} \rangle_{\beta_f} \\
\rho_f \\
\rho_f u_f \\
n \rho_f \Theta_f \langle m^{-1} \rangle_{\beta_f}
\end{array} \right) = \sum_{m \geq 1} \int_{{\mathbb R}^n} f_m(v) \left( \begin{array}{c} 
1 \\
m \\
mv \\
m |v-u_f|^2
\end{array} \right) \, dv, 
\label{eq:rhouthetabet_def}
\end{equation}
and we consider the following BGKME equation: 
\begin{equation}
\partial_t f_m^\varepsilon + v \cdot \nabla_x f_m^\varepsilon = \frac{1}{\varepsilon} \Big( \rho_{f^\varepsilon} \big( M_{u_{f^\varepsilon}, \Theta_{f^\varepsilon}, \beta_{f^\varepsilon}} \big)_m - f_m^\varepsilon \Big), \quad \forall m \geq 1. 
\label{eq:BGK}
\end{equation}
We introduce the Navier-Stokes mass-exchange model (NSME), of unknowns $(\rho, u, \Theta, \beta)$: ${\mathbb R}^n \times [0,T] \to [0,\infty) \times {\mathbb R}^n \times [0,\infty) \times {\mathcal S}$, as follows: 
\begin{eqnarray}
&&\hspace{-1cm}
\partial_t \big(\rho \langle m^{-1} \rangle_\beta \big) + \nabla_x \cdot \big( \rho \langle m^{-1} \rangle_\beta \, u \big) = \varepsilon \, \nabla_x \cdot ( \nu \, \nabla_x \chi ) , \label{eq:diff_popu} \\
&&\hspace{-1cm}
\partial_t \rho + \nabla_x \cdot (\rho u) = 0, \label{eq:diff_mass} \\
&&\hspace{-1cm}
\partial_t (\rho u) + \nabla_x \cdot (\rho u \otimes u) + \nabla_x \big( \rho \Theta \langle m^{-1} \rangle_\beta \big) = \varepsilon \, \nabla_x \cdot ( \mu \, \sigma(u) ), \label{eq:diff_mom} \\
&&\hspace{-1cm}
\partial_t \big( \rho |u|^2 + n \rho \Theta \langle m^{-1} \rangle_\beta \big) + \nabla_x \cdot \big( \rho |u|^2 u + (n+2) \rho \Theta \langle m^{-1} \rangle_\beta \,  u \big) \nonumber \\
&&\hspace{2cm}
= \varepsilon \, \nabla_x \cdot \big( (n+2) \, \nu \, \Theta \, \nabla_x \chi + 2 \kappa \, \nabla_x \Theta + 2 \mu \, \sigma(u) \, u \big), \label{eq:diff_energy}
\end{eqnarray}
where 
\begin{eqnarray}
\mu &=& \rho \Theta \langle m^{-1} \rangle_\beta, \label{eq:mu} \\
\kappa &=& \frac{n+2}{2} \, \rho \Theta \, \langle m^{-2} \rangle_\beta, \label{eq:kappa} \\
\nu &=&  \rho \Theta \, \big( \langle m^{-2} \rangle_\beta - \langle m^{-1} \rangle_\beta^2 \big), \label{eq:nu}
\end{eqnarray}
and where 
\begin{equation} 
\sigma(u) = \nabla_x u + (\nabla_x u)^T - \frac{2}{n} (\nabla_x \cdot u) \, \mathrm{I}_n, 
\label{eq:sigma_def}
\end{equation}
is the traceless rate-of-strain tensor (the exponent ``$T$'' standing for the matrix transpose operation) and 
\begin{equation} \chi = \log \left( \frac{\rho \Theta}{\sum_{m=1}^\infty \frac{m e^{\beta m}}{\gamma_m}} \right) = \log \left( (2 \pi)^{\frac{n}{2}} \frac{\rho \Theta^{\frac{n+2}{2}}}{Z(\beta,\Theta)} \right), 
\label{eq:chi_def}
\end{equation} 
is akin to a ``population potential''. The coefficients $\mu$, $\kappa$ and $\nu$ are the viscosity, heat conductivity and population diffusivity respectively. The dependence of $(\rho, u, \Theta, \beta)$ on $\varepsilon$ is omitted for simplicity. We first note the following 

\begin{lemma}[Positivity of $\nu$, $\kappa$, $\mu$] We have
\begin{equation}
\nu > 0,  \qquad \kappa >0,  \qquad \mu >0.  
\label{eq:nu_pos}
\end{equation}
\label{lem:nukappamu>0}
\end{lemma}

\noindent
\textbf{Proof.} The positivity of $\kappa$ and $\mu$ are obvious. We can express $\nu$ as 
$$ \nu = \rho \Theta \, \frac{\displaystyle \bigg( \sum_{m \geq 1} \frac{e^{\beta m}}{\gamma_m} \frac{1}{m} \bigg) \, \bigg( \sum_{m \geq 1} \frac{e^{\beta m}}{\gamma_m} m \bigg) - \bigg( \sum_{m \geq 1} \frac{e^{\beta m}}{\gamma_m} \bigg)^2} {\displaystyle \bigg( \sum_{m \geq 1} \frac{e^{\beta m}}{\gamma_m} \bigg)^2} =: \rho \Theta \, \frac{N}{D}, $$
with $D > 0$ and 
\begin{eqnarray*}
N&=& \sum_{m,p \geq 1} \frac{e^{\beta m}}{\gamma_m} \frac{e^{\beta p}}{\gamma_p}  \, \frac{p-m}{m} 
= \frac{1}{2} \sum_{m,p \geq 1} \frac{e^{\beta m}}{\gamma_m} \frac{e^{\beta p}}{\gamma_p} \, \Big( \, \frac{p-m}{m} + \frac{m-p}{p} \, \Big) \\
&=& \frac{1}{2} \sum_{m,p \geq 1} \frac{e^{\beta m}}{\gamma_m} \frac{e^{\beta p}}{\gamma_p} \, \frac{(p-m)^2}{mp}  > 0 , 
\end{eqnarray*}
which shows that $\nu >0$. \endproof

Next, we have the following theorem:

\begin{theorem}[NSME is a higher order approximation to BGKME]~

\noindent
Let $f^\varepsilon$ be a solution to  \eqref{eq:BGK} and define $(\rho_{f^\varepsilon}, u_{f^\varepsilon}, \Theta_{f^\varepsilon}, \beta_{f^\varepsilon})$ via \eqref{eq:rhouthetabet_def}. Then, $(\rho_{f^\varepsilon}, u_{f^\varepsilon}, \Theta_{f^\varepsilon}, \beta_{f^\varepsilon})$ satisfies the NSME system \eqref{eq:diff_popu}-\eqref{eq:diff_energy} up to terms of order ${\mathcal O}(\varepsilon^2)$.
\label{thm:CE}
\end{theorem}

\noindent
\textbf{Proof.} We will denote  $(\rho_{f^\varepsilon}, u_{f^\varepsilon}, \Theta_{f^\varepsilon}, \beta_{f^\varepsilon})$ simply by $(\rho, u, \Theta, \beta)$ and we show that $(\rho, u, \Theta, \beta)$  satisfies system \eqref{eq:diff_popu}-\eqref{eq:diff_energy} up to terms of order ${\mathcal O}(\varepsilon^2)$. We introduce $g^\varepsilon = (g^\varepsilon_m)_{m \geq 1}$ such that
\begin{equation} 
g^\varepsilon_m = \frac{1}{\varepsilon} \big( f^\varepsilon_m - \rho (M_{u, \Theta, \beta})_m \big), \quad \forall m \geq 1.  
\label{eq:geps_def}
\end{equation}
We also introduce the short-hand notation 
$$ D \equiv \partial_t + v \cdot \nabla_x. $$
By \eqref{eq:BGK} and iterating with \eqref{eq:geps_def}, we have 
$$ g^\varepsilon_m  = - D f^\varepsilon_m = - D \big( \rho (M_{u, \Theta, \beta})_m \big) + {\mathcal O}(\varepsilon), $$
and so 
\begin{eqnarray*} 
D f^\varepsilon_m &=& D \big( \rho (M_{u, \Theta, \beta})_m + \varepsilon D g^\varepsilon_m \big) \\
&=& D \Big( \rho (M_{u, \Theta, \beta})_m - \varepsilon D \big( \rho (M_{u, \Theta, \beta})_m \big) \Big) + {\mathcal O}(\varepsilon^2). 
\end{eqnarray*}
Hence, thanks to \eqref{eq:conservations}, we get
\begin{eqnarray}
&& \hspace{-1cm}
0 = \sum_{m \geq 1} \int_{{\mathbb R}^n} D f^\varepsilon_m  \left( \begin{array}{c} 
1 \\
m \\
mv \\
m |v|^2
\end{array} \right) \, dv  \nonumber \\
&& \hspace{-0.5cm}
= \sum_{m \geq 1} \int_{{\mathbb R}^n} D \Big( \rho (M_{u, \Theta, \beta})_m - \varepsilon D \big( \rho (M_{u, \Theta, \beta})_m \big) \Big) \left( \begin{array}{c} 
1 \\
m \\
mv \\
m |v|^2
\end{array} \right) \, dv + {\mathcal O}(\varepsilon^2).
\label{eq:moments=0} 
\end{eqnarray}

The next step is to compute $D ( \rho (M_{u, \Theta, \beta})_m )$ in terms of spatial derivatives of $(\rho, u, \Theta, \beta)$ only, up to terms of order ${\mathcal O}(\varepsilon)$. First, we notice that
$$ D \big( \rho (M_{u, \Theta, \beta})_m \big) = \rho (M_{u, \Theta, \beta})_m \, \Big\{ \, D (\log \rho)  + D (\log (M_{u, \Theta, \beta})_m) \, \Big\}. 
$$
From \eqref{eq:normalization}, we have 
\begin{equation}
\partial_\Theta \log Z (\beta, \Theta) = \frac{n}{2 \Theta}, \qquad
\partial_\beta \log Z (\beta, \Theta) = \langle m \rangle_\beta.
\label{eq:partial_Z}
\end{equation}
Then, with \eqref{eq:equilibria}, we find
\begin{eqnarray}
&&\hspace{-1cm}
D \big( \rho (M_{u, \Theta, \beta})_m \big) = \rho (M_{u, \Theta, \beta})_m \, \Big\{ \,  D (\log \rho)  + (m- \langle m \rangle_\beta) \, D \beta \nonumber \\ 
&&\hspace{5cm} 
+ \Big( \frac{m |v-u|^2}{\Theta} - n \Big) \,  \frac{D \Theta}{2 \Theta} + m \, \frac{v-u}{\Theta} \cdot Du \, \Big\}. 
\label{eq:DrhoM}
\end{eqnarray}

Now, we note that $(\rho, u, \Theta, \beta)$ satisfies the EME system in non-conservative form \eqref{eq:pop_NC}-\eqref{eq:temp_NC} with ${\mathcal O}(\varepsilon)$ terms at the right-hand side. We use this to replace the time derivatives in $D(\rho, u, \Theta, \beta)$ by spatial derivatives. We get
\begin{eqnarray*}
D \log \rho &=& - \nabla_x \cdot u + (v-u) \cdot \frac{\nabla_x \rho}{\rho} + {\mathcal O}(\varepsilon), \\
D \beta &=& (v-u) \cdot \nabla_x \beta + {\mathcal O}(\varepsilon), \\
D \Theta &=& (v-u) \cdot \nabla_x \Theta - \frac{2}{n} \Theta (\nabla_x \cdot u) + {\mathcal O}(\varepsilon), \\
Du &=& \big( (v-u) \cdot \nabla_x \big) u - \frac{1}{\rho} \nabla_x \big( \rho \Theta \langle m^{-1} \rangle_\beta \big)= {\mathcal O}(\varepsilon). 
\end{eqnarray*}
Inserting these relations into \eqref{eq:DrhoM} and using that 
\begin{equation} 
\frac{\partial \langle m^{-1} \rangle_\beta}{\partial \beta} = 1 -  \langle m^{-1} \rangle_\beta \,  \langle m \rangle_\beta, 
\label{eq:partial_beta_m-1}
\end{equation}
and 
$$ \frac{\nabla_x (\rho \Theta \langle m^{-1} \rangle_\beta)}{\rho} = \Theta \langle m^{-1} \rangle_\beta \frac{\nabla_x \rho}{\rho} + \langle m^{-1} \rangle_\beta \nabla_x \Theta + \Theta \big( 1 -  \langle m^{-1} \rangle_\beta \,  \langle m \rangle_\beta \big) \nabla_x \beta, $$
we get
\begin{eqnarray}
&&\hspace{-1cm}
D \big( \rho (M_{u, \Theta, \beta})_m \big) = {\mathcal E}_m + {\mathcal O}(\varepsilon), 
\label{eq:DrhoM_2}
\end{eqnarray}
with 
\begin{eqnarray}
&&\hspace{-1cm}
{\mathcal E}_m = \rho (M_{u, \Theta, \beta})_m \, \Big\{ \,  \frac{1}{\Theta} \, {\mathbb T}_m : \nabla_x u \nonumber \\
&&\hspace{3cm}
+ {\mathbf V}_m \cdot \Big[ \, \frac{\nabla_x \rho}{\rho} - \langle m \rangle_\beta \, \nabla_x \beta  \, \Big] 
+ {\mathbf W}_m \cdot \frac{\nabla_x \Theta}{2 \Theta} \, \Big\}, 
\label{eq:callEdef}
\end{eqnarray}
and
\begin{eqnarray}
{\mathbb T}_m &=&   m \Big( (v-u) \otimes (v-u) - \frac{|v-u|^2}{n} \, \mathrm{I}_n \Big),  \label{eq:Tdef} \\
{\mathbf V}_m &=& \big( \, 1 - m \langle m^{-1} \rangle_\beta \, \big) \, \, (v-u), \label{eq:boldVdef} \\
{\mathbf W}_m &=& \Big( \, m \, \frac{|v-u|^2}{\Theta} - n - 2 \, m \, \langle m^{-1} \rangle_\beta \, \Big) \, (v-u). 
\label{eq:boldWdef}
\end{eqnarray}
We note that ${\mathbb T}_m$ and $\nabla_x u$ are rank-2 tensor while ${\mathbf V}_m$ and ${\mathbf W}_m$ are vectors. The symbol~'$:$' denotes the contraction of two rank two tensors. We denote by ${\mathbb T} = ({\mathbb T}_m)_{m \geq 1}$ and similarly for ${\mathbf V}$ and ${\mathbf W}$. Let $\varphi$ denote any component of ${\mathbb T}$, ${\mathbf V}$ or ${\mathbf W}$. We can check that $\varphi$ satisfies
\begin{equation}
\sum_{m \geq 1} \int_{{\mathbb R}^n} (M_{u, \Theta, \beta})_m \, \, \varphi_m  \left( \begin{array}{c} 
1 \\
m \\
mv \\
m |v|^2
\end{array} \right) \, dv = 0, 
\label{eq:mom_Mphi=0}
\end{equation}
or equivalently that 
\begin{equation}
\sum_{m \geq 1} \int_{{\mathbb R}^n} (M_{u, \Theta, \beta})_m \, \, \varphi_m  \left( \begin{array}{c} 
1 \\
m \\
m (v-u) \\
m |v -u|^2
\end{array} \right) \, dv = 0. 
\label{eq:mom_Mphi=0_2}
\end{equation}
Indeed, this is a simple computation using \eqref{eq:moments_maxwellian_1} and \eqref{eq:moments_maxwellian_2} from Appendix \ref{app_remark_form}.

Then, we insert \eqref{eq:DrhoM_2} into \eqref{eq:moments=0} and get 
\begin{equation}
\sum_{m \geq 1} \int_{{\mathbb R}^n} \Big( D \big( \rho (M_{u, \Theta, \beta})_m \big) - \varepsilon D {\mathcal E}_m \Big) \left( \begin{array}{c} 
1 \\
m \\
mv \\
m |v|^2
\end{array} \right) \, dv = {\mathcal O}(\varepsilon^2).
\label{eq:moments=0_2} 
\end{equation}
We now compute the integrals. The first term $D (\rho (M_{u, \Theta, \beta})_m)$ gives rise to the left-hand side of Eqs. \eqref{eq:diff_popu}-\eqref{eq:diff_energy}. Indeed, the computations are the same as those made to show Theorem \ref{th:hydro_limit} and are not repeated. Thus, we focus on $ D {\mathcal E}_m = \partial_t {\mathcal E}_m + v \cdot \nabla_x {\mathcal E}_m$. We have 
$$
\sum_{m \geq 1} \int_{{\mathbb R}^n} \partial_t {\mathcal E}_m \left( \begin{array}{c} 
1 \\
m \\
mv \\
m |v|^2
\end{array} \right) \, dv = 
\partial_t \bigg( \sum_{m \geq 1} \int_{{\mathbb R}^n}  {\mathcal E}_m \left( \begin{array}{c} 
1 \\
m \\
mv \\
m |v|^2
\end{array} \right) \, dv \bigg) = 0, 
$$
thanks to \eqref{eq:mom_Mphi=0} and 
$$
\sum_{m \geq 1} \int_{{\mathbb R}^n} v \cdot \nabla_x {\mathcal E}_m \left( \begin{array}{c} 
1 \\
m \\
mv \\
m |v|^2
\end{array} \right) \, dv = 
\nabla_x \cdot \bigg( \sum_{m \geq 1} \int_{{\mathbb R}^n}  {\mathcal E}_m \left( \begin{array}{c} 
v \\
m v\\
m v \otimes v \\
m |v|^2 v
\end{array} \right) \, dv \bigg), 
$$
so that we are left to compute the moments of ${\mathcal E} = ({\mathcal E}_m)_{m \geq 1}$ appearing at the right hand side. We successively compute the different lines. 

\medskip
\noindent
\textbf{First line.}
We write 
\begin{eqnarray*}
\sum_{m \geq 1} \int_{{\mathbb R }^n} {\mathcal E}_m \, v \, dv &=& \sum_{m \geq 1} \int_{{\mathbb R }^n} {\mathcal E}_m \, (v-u) \, dv + u \sum_{m \geq 1} \int_{{\mathbb R }^n} {\mathcal E}_m \, \, dv \\
&=& \sum_{m \geq 1} \int_{{\mathbb R }^n} {\mathcal E}_m \, (v-u) \, dv, 
\end{eqnarray*}
by the first line of \eqref{eq:mom_Mphi=0}. Since ${\mathbb T}_m$ is even with respect to $(v-u)$, its contribution is $0$ by antisymmetry. By \eqref{eq:equilibria}, \eqref{eq:normalization} and \eqref{eq:moments_maxwellian_2} for $p=0$, we have 
\begin{eqnarray*}
\sum_{m \geq 1} \int_{{\mathbb R }^n}  \rho(M_{u, \Theta, \beta})_m \, (v-u) \otimes {\mathbf V}_m   \, dv &=& \frac{\rho}{Z(\beta, \Theta)} (2 \pi \Theta)^{\frac{n}{2}} \Theta  \sum_{m \geq 1} \frac{e^{\beta m}}{\gamma_m} \big( m^{-1} - \langle m^{-1}  \rangle_\beta \big) \, \mathrm{I}_n \\
&=& \rho \Theta \big( \langle m^{-2}  \rangle_\beta - (\langle m^{-1}  \rangle_\beta)^2 \big)  \, \mathrm{I}_n. 
\end{eqnarray*}
Similarly, with \eqref{eq:moments_maxwellian_2} for $p=0$ and $p=1$, we get
$$\sum_{m \geq 1} \int_{{\mathbb R }^n}  \rho (M_{u, \Theta, \beta})_m \, (v-u) \otimes {\mathbf W}_m  \, dv = 2 \rho \Theta \big( \langle m^{-2}  \rangle_\beta - (\langle m^{-1}  \rangle_\beta)^2 \big)  \, \mathrm{I}_n. 
$$
Collecting all these terms and using \eqref{eq:nu}, we get 
$$\sum_{m \geq 1} \int_{{\mathbb R}^n} {\mathcal E}_m \, v \, dv = \nu \Big[ \frac{\nabla_x \rho}{\rho} - \langle m \rangle_\beta \nabla_x \beta + \frac{\nabla_x \Theta}{\Theta} \Big] = \nu \big[ \nabla_x \log (\rho \Theta) - \langle m \rangle_\beta \nabla_x \beta \big]. $$
Now, we note that 
$$ \langle m \rangle_\beta \nabla_x \beta = \frac{\sum_{m=1}^\infty \frac{m^2 e^{\beta m}}{\gamma_m}}{\sum_{m=1}^\infty \frac{m e^{\beta m}}{\gamma_m}} \nabla_x \beta = \nabla_x \log \left( \sum_{m=1}^\infty \frac{m e^{\beta m}}{\gamma_m} \right). $$
Hence, using \eqref{eq:chi_def}, we get:
$$\sum_{m \geq 1} \int_{{\mathbb R}^n} {\mathcal E}_m \, v \, dv =  \nu \, \nabla_x  \log \left( \frac{\rho \Theta}{\sum_{m=1}^\infty \frac{m e^{\beta m}}{\gamma_m}} \right) = \nu \, \nabla_x \chi, $$
which yields the right-hand side of \eqref{eq:diff_popu}.

\medskip
\noindent
\textbf{Second line.}
The second line is just $0$ by the third line of \eqref{eq:mom_Mphi=0}, which leads to the right-hand side of \eqref{eq:diff_mass} being $0$.

\medskip
\noindent
\textbf{Third line.}
We note that 
\begin{eqnarray*}
\sum_{m \geq 1} \int_{{\mathbb R }^n} {\mathcal E}_m \, m v \otimes v \, dv &=& \sum_{m \geq 1} \int_{{\mathbb R }^n} {\mathcal E}_m \, m (v-u) \otimes (v-u) \, dv  \\
&&\hspace{1cm}
+ \sum_{m \geq 1} \int_{{\mathbb R }^n} {\mathcal E}_m \, \big( - m u \otimes u + m v \otimes u + m u \otimes v \big) dv \\
&=& \sum_{m \geq 1} \int_{{\mathbb R }^n} {\mathcal E}_m \, m (v-u) \otimes (v-u) \, dv , 
\end{eqnarray*}
by the second and third lines of \eqref{eq:mom_Mphi=0}. Since ${\mathbf V}_m$ and ${\mathbf W}_m$ are odd with respect to $v-u$, their contribution is $0$ by antisymmetry. Now, we use \eqref{eq:moments_maxwellian_3} and \eqref{eq:moments_maxwellian_2} with $p=1$ from Appendix \ref{app_remark_form} as well as \eqref{eq:normalization} and get: 
\begin{eqnarray*}
&&\hspace{-1cm}
\sum_{m \geq 1} \int_{{\mathbb R }^n}  \rho (M_{u, \Theta, \beta})_m \, {\mathbb T}_m : (\nabla_x u) \, m \, (v-u) \otimes (v-u) \, dv  \\
&&\hspace{0cm}
= \frac{\rho}{Z(\beta, \Theta)} (2 \pi \Theta)^{\frac{n}{2}} \Theta  \sum_{m \geq 1} \frac{e^{\beta m}}{\gamma_m} \Big( \nabla_x u + (\nabla_x u)^T +  (\nabla_x \cdot u) \mathrm{I}_n - \frac{n+2}{n} (\nabla_x \cdot u) \mathrm{I}_n \Big) \\
&&\hspace{0cm}
=  \rho \Theta  \langle m^{-1}  \rangle_\beta \, \sigma(u) = \mu \sigma(u), 
\end{eqnarray*}
where we have used \eqref{eq:mu} and \eqref{eq:sigma_def}. This leads to the right-hand side of \eqref{eq:diff_mom}.

\medskip
\noindent
\textbf{Fourth line.}
Using that 
$$ v |v|^2 = (v-u) |v-u|^2 + 2 \big( (v-u) \otimes (v-u) \big) u + |v-u|^2 u + v |u|^2 + 2 (u \cdot v) u - 2 |u|^2 u, $$
and the conservation identities \eqref{eq:mom_Mphi=0}, \eqref{eq:mom_Mphi=0_2}, we get
\begin{eqnarray*}
\sum_{m \geq 1} \int_{{\mathbb R }^n} {\mathcal E}_m \, m \, v \, |v|^2 \, dv &=& \sum_{m \geq 1} \int_{{\mathbb R }^n} {\mathcal E}_m \, m \, (v-u) \, |v-u|^2 \, dv  \\
&&\hspace{1cm}
+ 2 \bigg( \sum_{m \geq 1} \int_{{\mathbb R }^n} {\mathcal E}_m \, m (v-u) \otimes (v-u) \, dv \bigg) u.
\end{eqnarray*}
The second term has already been computed when dealing with the third line and gives:
$$ 2 \bigg( \sum_{m \geq 1} \int_{{\mathbb R }^n} {\mathcal E}_m \, m (v-u) \otimes (v-u) \, dv \bigg) u = 2 \mu \, \sigma(u) \, u . $$
Then, we focus on the first term and notice that the contribution of ${\mathbb T}_m$ is $0$ by antisymmetry. We have, using \eqref{eq:moments_maxwellian_2} with $p=1$ and \eqref{eq:normalization}: 
\begin{eqnarray*}
&&\hspace{-1cm}
\sum_{m \geq 1} \int_{{\mathbb R }^n} \rho (M_{u, \Theta, \beta})_m \, m \, {\mathbf V}_m \otimes (v-u) \, |v-u|^2 \, dv \\ 
&& \hspace{2cm}
= \frac{\rho}{Z(\beta, \Theta)} (2 \pi \Theta)^{\frac{n}{2}} \Theta^2 (n+2)   \sum_{m \geq 1} \frac{e^{\beta m}}{\gamma_m} \Big( m^{-1} - \langle m^{-1} \rangle_\beta \Big) \, \mathrm{I}_n \\
&& \hspace{2cm}
= (n+2) \rho \Theta^2 \Big( \langle m^{-2} \rangle_\beta - (\langle m^{-1} \rangle_\beta)^2 \Big) \, \mathrm{I}_n. 
\end{eqnarray*}
Similarly, thanks to \eqref{eq:moments_maxwellian_2} with $p=1$ and $p=2$ and \eqref{eq:normalization}: 
\begin{eqnarray*}
&&\hspace{-1cm}
\sum_{m \geq 1} \int_{{\mathbb R }^n} \rho (M_{u, \Theta, \beta})_m \, m \, {\mathbf W}_m \otimes (v-u) \, |v-u|^2 \, dv \\ 
&& \hspace{-0.8cm}
= \frac{\rho}{Z(\beta, \Theta)} (2 \pi \Theta)^{\frac{n}{2}} \Theta^2 \sum_{m \geq 1} \frac{e^{\beta m}}{\gamma_m} \frac{1}{m} \Big( (n+2)(n+4) - n (n+2) - 2 (n+2) m \langle m^{-1} \rangle_\beta \Big) \, \mathrm{I}_n \\
&& \hspace{-0.8cm}
= 2 (n+2) \rho \Theta^2 \Big( 2 \langle m^{-2} \rangle_\beta - (\langle m^{-1} \rangle_\beta)^2 \Big) \, \mathrm{I}_n. 
\end{eqnarray*}
Collecting all these terms and using the same algebra as for the first line, as well as \eqref{eq:kappa} leads to 
\begin{eqnarray*}
&&\hspace{-1cm}
\sum_{m \geq 1} \int_{{\mathbb R }^n} \rho (M_{u, \Theta, \beta})_m \, m \, {\mathbf W}_m \otimes (v-u) \, |v-u|^2 \, dv \\ 
&& \hspace{0cm}
= (n+2) \rho \Theta^2 \, \Big( \langle m^{-2} \rangle_\beta - (\langle m^{-1} \rangle_\beta)^2 \Big) \, \Big[ \frac{\nabla_x \rho}{\rho} - \langle m \rangle_\beta \nabla_x \beta + \frac{\nabla_x \Theta}{\Theta} \Big] \\
&& \hspace{9cm}
+ (n+2) \rho \Theta \, \langle m^{-2} \rangle_\beta \, \nabla_x \Theta \\
&& \hspace{0cm}
= (n+2) \Theta \nu \, \nabla_x \chi + 2 \kappa \, \nabla_x \Theta, 
\end{eqnarray*}
which yields the right-hand side of \eqref{eq:diff_energy} and ends the proof. \endproof

\setcounter{equation}{0}
\section{Entropy and thermodynamics}
\label{sec_entropy}

\subsection{Kinetic entropy}
\label{subsec_entropy}

\begin{proposition}[Kinetic entropy]~

\noindent
Let $f=(f_m)_{m \geq 1}$ be a solution of the BME equation \eqref{eq:Boltz_eq} or the BGKME equation~\eqref{eq:BGK}. Define the kinetic entropy by
\begin{equation}
S(f) = \sum_{m \geq 1} \int_{{\mathbb R}^n} f_m \, \Big[ \log \Big( \frac{\gamma_m f_m}{m^{\frac{n}{2}}} \Big) -1 \Big] \, dv, 
\label{eq:entropy} 
\end{equation} 
and the kinetic entropy flux by 
\begin{equation}
\phi(f) = \sum_{m \geq 1} \int_{{\mathbb R}^n} f_m \, \Big[ \log \Big( \frac{\gamma_m f_m}{m^{\frac{n}{2}}} \Big) -1 \Big] \, v \, dv.  
\label{eq:entropy_flux} 
\end{equation}
Then, we have 
$$ \frac{\partial S}{\partial t} + \nabla_x \cdot \phi \leq 0. $$
\label{prop:entropy}
\end{proposition}

\noindent
\textbf{Proof.} (i) for solutions of \eqref{eq:Boltz_eq}: we multiply \eqref{eq:Boltz_eq} by $\log (\frac{\gamma_m f_m}{m^{\frac{n}{2}}})$ and use \eqref{eq:entropy_decay} as well as the fact that $\frac{d}{df} (f(\log (cf) -1)) = \log (cf)$, for any constant $c>0$. 

\noindent
(ii) for solutions of \eqref{eq:BGK}: we proceed in the same way but must prove an analog of \eqref{eq:entropy_decay} for the BGKME operator. This is classical. We remark that 
\begin{equation} 
\log \Big( \frac{\gamma_m \rho (M_{u, \Theta, \beta})_m}{m^{\frac{n}{2}}} \Big) = \log \rho - \log Z + \beta m - \frac{m|v-u|^2}{2 \Theta},
\label{eq:log_equi}
\end{equation}
is a linear combination of $1$, $m$, $mv$ and $m|v|^2$, so that by the definition \eqref{eq:rhouthetabet_def} of $(\rho_f,u_f,\Theta_f,\beta_f)$, we have 
$$ \sum_{m \geq 1} \int_{{\mathbb R}^n} Q(f)_m \, \log \Big( \frac{\gamma_m \rho (M_{u, \Theta, \beta})_m}{m^{\frac{n}{2}}} \Big) \, dv = 0. $$
Therefore,
\begin{eqnarray*}
&&\hspace{-1cm}
 \sum_{m \geq 1} \int_{{\mathbb R}^n} Q(f)_m \, \log \Big( \frac{\gamma_m f_m}{m^{\frac{n}{2}}} \Big) \, dv \\
&&\hspace{0cm}
= - \sum_{m \geq 1} \int_{{\mathbb R}^n} \Big( f_m - \rho_f (M_{u, \Theta, \beta})_m \Big) \, \Big[ \log \Big( \frac{\gamma_m f_m}{m^{\frac{n}{2}}} \Big) - \log \Big( \frac{\gamma_m \rho (M_{u, \Theta, \beta})_m}{m^{\frac{n}{2}}} \Big) \Big] \, dv \\ 
&&\hspace{0cm}
= - \sum_{m \geq 1} \int_{{\mathbb R}^n} \Big( f_m - \rho_f (M_{u, \Theta, \beta})_m \Big) \, \Big[ \log  f_m - \log \big( \rho (M_{u, \Theta, \beta})_m \big) \Big] \, dv \leq 0, 
\end{eqnarray*}
by the fact that $\log$ is an increasing function. \endproof

\begin{proposition}[Entropy and entropy flux at local equilibrium]~

\noindent
We have 
\begin{eqnarray}
S(\rho M_{u, \Theta, \beta}) &=& \rho \Big( \langle m^{-1} \rangle_\beta \, \big( \log \rho - \log Z - 1 - \frac{n}{2} \big) + \beta \Big), \label{eq:entrop_equi} \\
\phi(\rho M_{u, \Theta, \beta}) &=& S(\rho M_{u, \Theta, \beta}) \, u. 
\label{eq:entrop_flux_equi}
\end{eqnarray}
\end{proposition}

\noindent
\textbf{Proof.} With \eqref{eq:log_equi}, we get 
$$ S(\rho M_{u, \Theta, \beta}) = \frac{\rho}{Z} \sum_{m \geq 1} \int_{{\mathbb R}^n}
\frac{m^{\frac{n}{2}} e^{\beta m}}{\gamma_m} \Big( \big[ \log \rho - \log Z - 1 \big] + \beta m - \frac{m|v-u|^2}{2 \Theta} \Big) \, e^{-\frac{m|v-u|^2}{2 \Theta}} \, dv.  $$
Thanks to \eqref{eq:moments_maxwellian_1} and \eqref{eq:normalization}, this formula leads to \eqref{eq:entrop_equi}. Then, we have 
\begin{eqnarray} \phi(\rho M_{u, \Theta, \beta}) &=& \sum_{m \geq 1} \int_{{\mathbb R}^n} (\rho M_{u, \Theta, \beta})_m \, \Big[ \log \Big( \frac{\gamma_m (\rho M_{u, \Theta, \beta})_m}{m^{\frac{n}{2}}} \Big) -1 \Big] \, v \, dv \label{eq:entrop_flux_alt} \\  
&=& \Big( \sum_{m \geq 1} \int_{{\mathbb R}^n} (\rho M_{u, \Theta, \beta})_m \, \Big[ \log \Big( \frac{\gamma_m (\rho M_{u, \Theta, \beta})_m}{m^{\frac{n}{2}}} \Big) -1 \Big] \, dv \Big) \, u, \nonumber
\end{eqnarray}
by antisymmetry, which leads to \eqref{eq:entrop_flux_equi}. \endproof

\subsection{Thermodynamic entropy}
\label{subsec_thermo}

The goal of this section is to introduce the thermodynamic entropy, to prove that it is equal to the kinetic entropy $S$ and that it is a convex function of the fluid moments (or conservative variables). We use the framework developed in \cite{levermore1996moment}. We introduce the following notations. Let 
\begin{equation}
{\mathcal A} = \left( \begin{array}{c} D \\ A \\ B \\ C \end{array} \right), 
\label{eq:calA_def}
\end{equation}
be the vector of parameters of the Maxwellian as written in \eqref{eq:maxwell_entropic_var}, with $A$, $B$, $C$ in ${\mathbb R}$ and $D$ in ${\mathbb R}^n$. Note that we list the components of ${\mathcal A}$ in the order depicted in \eqref{eq:calA_def}. We will denote the components of ${\mathcal A}$ by greek exponents, namely ${\mathcal A} = ({\mathcal A}^\alpha)_{\alpha = 1, \ldots, n+3}$, while we will keep latin indices $i=1, \ldots, n$ for spatial coordinates. Specifically, 
$$ {\mathcal A}^\alpha = D_\alpha, \quad \forall \alpha = 1, \ldots, n, \qquad {\mathcal A}^{n+1} = A, \qquad {\mathcal A}^{n+2} = B, \qquad {\mathcal A}^{n+3} = C. $$ 
${\mathcal A}$ is called the vector of entropic (or intensive) variables. 

Likewise, we introduce ${\boldsymbol \mu} = ({\boldsymbol \mu}_m)_{m \geq 1}$, where, for each $m$, ${\boldsymbol \mu}_m$ is the vector of velocity moments involved in the definition of the equilibrium moments \eqref{eq:moments_equi}, namely
$$ {\boldsymbol \mu}_m = \left( \begin{array}{c} mv \\ 1 \\ m \\ m|v|^2 \end{array} \right), \quad \textrm{ or equivalently } \quad {\boldsymbol \mu} = \left( \begin{array}{c} \mathbf{mv} \\ \mathbf{1} \\ \mathbf{m} \\ \mathbf{m|v|^2}  \end{array} \right). $$
Note that we adopt the same ordering as for the entropic variable ${\mathcal A}$. Thus, we have 
$$ {\boldsymbol \mu}^\alpha = \mathbf{mv}_\alpha, \quad \forall \alpha = 1, \ldots, n, \qquad {\boldsymbol \mu}^{n+1} = \mathbf{1}, \qquad {\boldsymbol \mu}^{n+2} = \mathbf{m}, \qquad {\boldsymbol \mu}^{n+3} = \mathbf{m|v|^2}. $$

Finally, denote by ${\mathcal M}$ the vector of equilibrium moments given by \eqref{eq:moments_equi}, i.e. 
\begin{equation} 
{\mathcal M} = \left( \begin{array}{c} P \\ {\mathcal N} \\ \rho \\ E \end{array} \right) =:\left( \begin{array}{c} \rho u \\ \rho \langle m^{-1} \rangle_\beta \\ \rho \\ \rho (|u|^2 + n \Theta \langle m^{-1} \rangle_\beta)  \end{array} \right). 
\label{eq:conserv_var_express}
\end{equation}
${\mathcal M}$ is called the vector of conservative (or extensive) variables. Its components are the momentum $P$, the number of particles (or population) ${\mathcal N}$ the mass $\rho$ and the total energy~$E$. Here again, we have changed the ordering of the components \eqref{eq:moments_equi} to fit the ordering of ${\mathcal A}$.  Hence, 
\begin{eqnarray*} {\mathcal M}^\alpha &=& \rho u_\alpha, \quad \forall \alpha = 1, \ldots, n, \\
{\mathcal M}^{n+1} &=& \rho \langle m^{-1} \rangle_\beta, \qquad  {\mathcal M}^{n+2} = \rho, \qquad  {\mathcal M}^{n+3} = \rho (|u|^2 + n \Theta \langle m^{-1} \rangle_\beta). 
\end{eqnarray*}
There is a one-to-one onto correspondence between ${\mathcal A}$ and ${\mathcal M}$. We can pass from ${\mathcal A}$ to~${\mathcal M}$ using \eqref{eq:map_entro2cons}, \eqref{eq:rel_rho_alpha} and these relations can be easily inverted. Below, we make the structure of this map more precise. 

In the thermodynamic framework, the entropy is the Legendre transform of the Massieu-Planck Potential \cite{balian1991microphysics} which we will denote by $\Sigma$. It is defined as a function of the entropic variable ${\mathcal A}$ as follows: 
$$ \Sigma({\mathcal A}) = \sum_{m \geq 1} \int_{{\mathbb R}^n} \frac{m^{\frac{n}{2}}}{\gamma_m} e^{{\boldsymbol \mu}_m(v) \bullet  {\mathcal A}} \,  dv, $$
where the symbol $\bullet$ is the Euclidean inner product in ${\mathbb R}^{n+3}$, which we distinguish from the inner product in ${\mathbb R}^n$ denoted by the standard symbol ``$\cdot$''. Given that 
$$ {\boldsymbol \mu}_m(v) \bullet  {\mathcal A} = mv \cdot D + A + mB + m|v|^2 C , $$ 
and owing to the relation \eqref{eq:maxwell_entropic_var}, we have 
\begin{equation} 
\frac{m^{\frac{n}{2}}}{\gamma_m} e^{{\boldsymbol \mu}_m(v) \bullet  {\mathcal A}} = \rho (M_{u,\Theta, \beta})_m, 
\label{eq:Maxw_in_entrop_var}
\end{equation}
with $(\rho, u,\Theta, \beta)$ related to ${\mathcal A}$ through \eqref{eq:map_entro2cons}, \eqref{eq:rel_rho_alpha}, so that 
\begin{equation} 
\Sigma({\mathcal A}) = \rho \langle m^{-1} \rangle_\beta. 
\label{eq:express_Sigma}
\end{equation} 
But we stress that $\rho \langle m^{-1} \rangle_\beta$ can only be identified with the Massieu-Planck potential if considered as a function of ${\mathcal A}$. 

Next, we note that 
$$ \nabla_{{\mathcal A}} \Sigma({\mathcal A}) = 
\sum_{m \geq 1} \int_{{\mathbb R}^n} \frac{m^{\frac{n}{2}}}{\gamma_m} e^{{\boldsymbol \mu}_m(v) \bullet  {\mathcal A}} \, {\boldsymbol \mu}_m(v) \, dv 
= \sum_{m \geq 1} \int_{{\mathbb R}^n} \rho (M_{u,\Theta, \beta})_m \, {\boldsymbol \mu}_m(v) \, dv = {\mathcal M}, $$
by \eqref{eq:moments_equi}. Introducing a vector-valued potential $\Phi$ by 
\begin{equation} 
\Phi({\mathcal A}) = \sum_{m \geq 1} \int_{{\mathbb R}^n} \frac{m^{\frac{n}{2}}}{\gamma_m} e^{{\boldsymbol \mu}_m(v) \bullet  {\mathcal A}} \, v \, dv, 
\label{eq:Big_Phi_def}
\end{equation}
we likewise see that $ \nabla_{{\mathcal A}} \Phi_i ({\mathcal A})$ is the flux in the direction $x_i$ of the Euler system \eqref{eq:hydro_limit}, so that the latter can be written as a system for the intensive variables ${\mathcal A}(x,t)$ as 
\begin{equation} 
\partial_t \Big( \nabla_{{\mathcal A}} \Sigma  \big( {\mathcal A} (x,t) \big)  \Big) + \sum_{i=1}^n \partial_{x_i} \Big( \nabla_{{\mathcal A}} \Phi_i  \big( {\mathcal A} (x,t) \big) \Big) = 0, 
\label{eq:EME_entrovar}
\end{equation}
or equivalently (for smooth solutions), as
\begin{equation}
\nabla_{\mathcal A}^2 \Sigma  \big( {\mathcal A} (x,t) \big) \, 
\partial_t {\mathcal A} (x,t)  + \sum_{i=1}^n \nabla_{\mathcal A}^2 \Phi_i \big( {\mathcal A} (x,t) \big) \, \partial_{x_i} {\mathcal A} (x,t) = 0, 
\label{eq:Euler_symmetrizable}
\end{equation}
We have 
\begin{equation} 
\nabla^2_{{\mathcal A}} \Sigma({\mathcal A}) = 
\sum_{m \geq 1} \int_{{\mathbb R}^n} \frac{m^{\frac{n}{2}}}{\gamma_m} e^{{\boldsymbol \mu}_m(v) \bullet  {\mathcal A}} \, {\boldsymbol \mu}_m(v) {\boldsymbol \otimes} {\boldsymbol \mu}_m(v) \, dv, 
\label{eq:nabla2_Sigma}
\end{equation}
where the symbol ${\boldsymbol \otimes}$ stands for the tensor product in ${\mathbb R}^{n+3}$ (while ``$\otimes$'' stands for the tensor product in ${\mathbb R}^n$). This shows that $\nabla^2_{{\mathcal A}} \Sigma({\mathcal A})$ is a symmetric matrix. A similar observation can be made for $\nabla^2_{{\mathcal A}} \Phi_i ({\mathcal A})$ for all $i=1, \ldots, n$. We immediately see that $\nabla^2_{{\mathcal A}} \Sigma({\mathcal A})$ is a positive matrix, as for any vector $\Xi \in {\mathbb R}^{n+3}$, we have 
$$ \Xi^T \nabla^2_{{\mathcal A}} \Sigma({\mathcal A}) \Xi = \sum_{m \geq 1} \int_{{\mathbb R}^n} \frac{m^{\frac{n}{2}}}{\gamma_m} e^{{\boldsymbol \mu}_m(v) \bullet  {\mathcal A}} \, ({\boldsymbol \mu}_m(v) \cdot \Xi)^2  \, dv \geq 0. $$
Below, we show that $\nabla^2_{{\mathcal A}} \Sigma({\mathcal A})$ is positive-definite which makes System \eqref{eq:Euler_symmetrizable} a symmetrizable hyperbolic system of equations for ${\mathcal A}$. 

\begin{proposition}[Strict convexity of the Massieu-Planck potential]~

\noindent
(i) Let $\rho$ and $\Theta$ be positive. Then, 
$\nabla^2_{{\mathcal A}} \Sigma({\mathcal A})$ is a positive-definite symmetric matrix. 

\noindent
(ii) $\Sigma$ is a strictly convex function of ${\mathcal A}$ in the domain of ${\mathbb A} \subset {\mathbb R}^{n+3}$ defined by 
\begin{equation} 
{\mathbb A} = \bigg\{ (D,A,B,C) \in {\mathbb R}^n \times {\mathbb R} \times {\mathbb R} \times {\mathbb R} \quad \Big| \quad C <0, \quad B - \frac{|D|^2}{4C} \in {\mathcal S} \bigg\}, 
\label{eq:boldA_def}
\end{equation}
where ${\mathcal S}$ is the set given by \eqref{eq:def_S}. 
\label{prop:positive_definite}
\end{proposition}

\noindent
\textbf{Proof.} (i) 
We write explicitly:
$$ {\boldsymbol \mu}_m(v) {\boldsymbol \otimes} {\boldsymbol \mu}_m(v) = \left( \begin{array}{cccc}
m^2 v \otimes v & m v & m^2 v & m^2 |v|^2 v\\
m v^T & 1 & m & m|v|^2  \\
m^2 v^T & m & m^2 & m^2 |v|^2\\
m^2 |v|^2 v^T &  m |v|^2 & m^2 |v|^2 & m^2 |v|^4
\end{array}
\right) $$
where the first column stands for $n$ columns, the first line for $n$ lines, and the upper left block element for an $n \times n$ block.  Inserting this into \eqref{eq:nabla2_Sigma} and using \eqref{eq:Maxw_in_entrop_var}, we realize that most of the entries of the resulting matrix are easily computed from previous calculations. The only one that deserves some further inspection is the lower right term because it is a fourth order velocity moment which has not been encountered before. We have 
\begin{eqnarray*} 
|v|^4 &=& \big( |v-u|^2 + |u|^2 + 2 (v-u).u \big)^2 \\
&=& |v-u|^4 + |u|^4 + 2 |v-u|^2 |u|^2 + 4 \big( (v-u) \cdot u \big)^2 + \textrm{ even terms in } (v-u). 
\end{eqnarray*}
Now, the corresponding integral can be computed by means of \eqref{eq:moments_maxwellian_1}, \eqref{eq:moments_maxwellian_2}. We finally get
\begin{eqnarray*}
&&\hspace{-1cm}
\rho^{-1} \nabla^2_{{\mathcal A}} \Sigma({\mathcal A})  \\
&&\hspace{0.cm}
= \left( \begin{array}{cccc}
\langle m \rangle u \otimes u + \Theta \mathrm{I}_n & u & \langle m \rangle u & \big( (n+2) \Theta + \langle m \rangle |u|^2 \big) u \\
* & \langle m^{-1} \rangle & 1 & |u|^2 + n \langle m^{-1} \rangle \Theta \\
* & * & \langle m \rangle & |u|^2 \langle m \rangle + n \Theta \\
* & * & * & (n+2) \Theta ( n \Theta \langle m^{-1} \rangle + 2 |u|^2 ) + \langle m \rangle |u|^4
\end{array}
\right),  
\end{eqnarray*}
where we have omitted the subscript $\beta$ to the averages $\langle m^k \rangle$ and we have displayed only the upper triangular part of the matrix, owing to its symmetry. Now, let $\Xi = (\zeta,\varphi, \xi, \eta)$ with $\zeta \in {\mathbb R}^n$ and $\varphi$, $\xi$, $\eta$ in ${\mathbb R}$. 

\medskip
\noindent {\em Assume $u \not = 0$.} We define 
$$ \zeta_\parallel = \frac{u \cdot \zeta}{|u|}, \quad \zeta_\bot = \zeta - \zeta_\parallel \, \frac{u}{|u|}. $$
We remark that $\rho^{-1} \Xi^T \nabla^2_{{\mathcal A}} \Sigma({\mathcal A}) \Xi$ is a quadratic form in the variables $(|\zeta_\bot|, \zeta_\parallel, \varphi, \xi, \eta, )$ whose matrix ${\mathbb S}$ is given by 
\begin{eqnarray*}
&&\hspace{-0.5cm}
{\mathbb S} =  
\left( \begin{array}{ccccc}
\Theta & 0 & 0 & 0 & 0 \\
* & \langle m \rangle |u|^2 + \Theta & |u| & \langle m \rangle |u| & \big( (n+2) \Theta + \langle m \rangle |u|^2 \big) |u|  \\
* & * & \langle m^{-1} \rangle & 1 & |u|^2 + n \langle m^{-1} \rangle \Theta \\
* & * & * & \langle m \rangle & |u|^2 \langle m \rangle + n \Theta \\
* & * & * & * & (n+2) \Theta ( n \Theta \langle m^{-1} \rangle + 2 |u|^2 ) + \langle m \rangle |u|^4
\end{array}
\right), 
\end{eqnarray*}

Provided that $\rho >0$, showing that $\nabla^2_{{\mathcal A}} \Sigma({\mathcal A}) $ is positive-definite is equivalent to showing that ${\mathbb S}$ is positive-definite. To show this, we apply Sylvester's criterion which says that ${\mathbb S}$ is positive-definite if and only if all its leading principal minors are positive. If  ${\mathbb S} = ({\mathbb S}_{ij})_{i, j = 1}^{5}$, its leading principal minors are $D_k = \det {\mathbb S}_k$ for $k=1, \ldots, 5$ where ${\mathbb S}_k = ({\mathbb S}_{ij})_{i, j = 1}^{k}$. We compute successively 
\begin{eqnarray}
D_1 &=& \Theta, \nonumber\\
D_2 &=& \Theta \, (\langle m \rangle |u|^2 + \Theta), \label{eq:D2}\\
D_3 &=& \Theta \, \big[ (\langle m^{-1} \rangle \langle m \rangle - 1) |u|^2 + \Theta \langle m^{-1} \rangle \big], \label{eq:D3}\\
D_4 &=& \Theta^2 \, (\langle m^{-1} \rangle \langle m \rangle - 1), \label{eq:D4}\\
D_5 &=& 2 n \Theta^4 \, \langle m^{-1} \, \rangle (\langle m^{-1} \rangle \langle m \rangle - 1). \label{eq:D5}  
\end{eqnarray}
We have $\langle m^{-1} \rangle \langle m \rangle - 1 >0$ by a similar proof to that of Lemma \ref{lem:nukappamu>0}. If follows that all determinants $D_1, \ldots, D_5$ are positive.  

\medskip
\noindent {\em Assume now that $u = 0$.} Then, $\rho^{-1} \Xi^T \nabla^2_{{\mathcal A}} \Sigma({\mathcal A}) \Xi$ is a quadratic form in the variables $(|\zeta|, \varphi, \xi, \eta)$ whose matrix $\tilde {\mathbb S}$ is given by 
\begin{eqnarray*}
&&\hspace{-0.5cm}
\tilde {\mathbb S} = 
\left( \begin{array}{cccc}
\Theta  & 0 & 0  &  0\\
* & \langle m^{-1} \rangle & 1 &  n \langle m^{-1} \rangle \Theta \\
* & * & \langle m \rangle &  n \Theta \\
* &   * & *  & n (n+2) \Theta^2 \langle m^{-1} \rangle \\
\end{array}
\right),
\end{eqnarray*}
We immediately see that its four principal minors $D'_i$, $i=1, \ldots, 4$ are such that $D'_i = D_{i+1}/\Theta$, where we have made $|u|=0$ in the formulas \eqref{eq:D2} to \eqref{eq:D5} for the $D_{i+1}$. It follows that the $D'_i$ are all positive for $i=1, \ldots, 4$.  

\medskip
\noindent {\em In all cases,} this shows that $\nabla^2_{{\mathcal A}} \Sigma({\mathcal A})$ is a positive-definite symmetric matrix. 

\medskip
\noindent
(ii) By (i), $ \Sigma$ is a strictly convex function of ${\mathcal A}$, for ${\mathcal A}$ being such that $\rho >0$ and $\Theta >0$. By the proof of Prop \ref{prop:equilibria}, it is immediate that this domain is the set ${\mathbb A}$ given by \eqref{eq:boldA_def}. This ends the proof of Proposition \ref{prop:positive_definite}. \endproof

Since $\Sigma$ is strictly convex on ${\mathbb A}$ the map $\nabla_{\mathcal A} \Sigma$: ${\mathbb A} \to {\mathbb R}^{n+3}$, ${\mathcal A} \mapsto {\mathcal M} = \nabla_{\mathcal A}\Sigma ({\mathcal A})$  is a local diffeomorphism. It is also straightforward to see that $\nabla_{\mathcal A} \Sigma$ is a one-to-one onto map ${\mathbb A} \to {\mathbb M}$, where 
\begin{equation} 
{\mathbb M} = \bigg\{ (P,{\mathcal N},\rho,E) \in {\mathbb R}^n \times {\mathbb R} \times {\mathbb R} \times {\mathbb R}  \quad \Big| \quad {\mathcal N} >0, \quad \rho >0, \quad E - \frac{|P|^2}{\rho} >0 \bigg\}. 
\label{eq:boldM_def}
\end{equation}
Thus, $\nabla_{\mathcal A} \Sigma$: ${\mathbb A} \to {\mathbb M}$ is a global diffeomorphism and we denote by $(\nabla_{\mathcal A} \Sigma)^{-1}$ its inverse. Since $\Sigma$ is strictly convex, we can define its Legendre transform 
\begin{equation} 
S({\mathcal M}) = {\mathcal A} \bullet {\mathcal M} - \Sigma ({\mathcal A}) \quad \textrm{with} \quad {\mathcal A} \quad \textrm{such that} \quad \nabla_{\mathcal A} \Sigma({\mathcal A}) =  {\mathcal M}, 
\label{eq:legendre_transform}
\end{equation}
which is called the thermodynamic entropy. Equivalently, we can write
$$ S({\mathcal M}) = \big( (\nabla_{\mathcal A} \Sigma)^{-1} ({\mathcal M}) \big) \bullet {\mathcal M} - \Sigma \big( (\nabla_{\mathcal A} \Sigma)^{-1} ({\mathcal M}) \big). $$
It is a classical fact, easy to check, that 
\begin{equation} 
\nabla_{\mathcal M} S ({\mathcal M})  = (\nabla_{\mathcal A} \Sigma)^{-1}({\mathcal M}) = (\nabla_{\mathcal A} \Sigma)^{-1} \big( \nabla_{\mathcal A} \Sigma({\mathcal A})\big)  = {\mathcal A} , 
\label{eq:leg_transf_deriv}
\end{equation}
i.e. the derivatives of $S$ and $\Sigma$ are inverse maps one to each other. By differentiating~\eqref{eq:leg_transf_deriv}, it follows that the following matrix equality holds:  
\begin{equation} 
\nabla^2_{\mathcal M} S ({\mathcal M}) = \Big(  \nabla^2_{\mathcal A} \Sigma \big( \nabla_{\mathcal M} S  ({\mathcal M}) \big)  \Big)^{-1}. 
\label{eq:leg_transf_sec_deriv}
\end{equation}
Since $\nabla^2_{\mathcal A} \Sigma ({\mathcal A})$ is positive-definite for all  ${\mathcal A} \in {\mathbb A}$, it follows that $\nabla^2_{\mathcal M} S ({\mathcal M})$ is positive-definite for all ${\mathcal M} \in {\mathbb M}$. Hence, $S$ is a strictly convex function of ${\mathcal M}$. The following statement establishes that $S$ given by \eqref{eq:legendre_transform} coincides with the fluid entropy \eqref{eq:entrop_equi}. 

\begin{proposition}[Strict convexity of the entropy]~

\noindent
(i) The thermodynamic entropy given by \eqref{eq:legendre_transform} and the kinetic entropy \eqref{eq:entrop_equi} are equal. 

\medskip
\noindent
(ii) The thermodynamic entropy $S$ is a strictly convex function of the conservative variables ${\mathcal M}$ for all ${\mathcal M} \in {\mathbb M}$. Its second derivative $\nabla^2_{\mathcal M} S ({\mathcal M})$ is positive-definite for all ${\mathcal M} \in {\mathbb M}$.
\label{prop:entropy_convex}
\end{proposition}

\noindent
\textbf{Proof.} Statement (ii) follows from (i) and the previous discussion. We focus on (i). We compute the Legendre transform of $\Sigma$ from \eqref{eq:legendre_transform}. We have 
$$S({\mathcal M}) = P \cdot D + {\mathcal N} A + \rho B + E C  - \Sigma(A). $$
Using \eqref{eq:conserv_var_express} to express ${\mathcal N}$, $\rho$, $E$, $P$ on the one hand, \eqref{eq:map_entro2cons}, \eqref{eq:rel_rho_alpha} to express $A$, $B$, $C$, $D$ on the other hand, and finally using \eqref{eq:express_Sigma}, we find
\begin{eqnarray*}
S({\mathcal M}) &=& \rho \langle m^{-1} \rangle_\beta (\log \rho - \log Z) + \rho \Big( \beta - \frac{|u|^2}{2 \Theta} \Big) \\
&& + \big( \rho |u|^2 + n \rho \langle m^{-1} \rangle_\beta \Theta \big) \Big( - \frac{1}{2 \Theta} \Big) + \rho u \cdot \Big( \frac{u}{\Theta} \Big) - \rho \langle m^{-1} \rangle_\beta \\
&=& \rho \langle m^{-1} \rangle_\beta \Big( \log \rho - \log Z - 1 - \frac{n}{2} \Big) + \rho \beta, 
\end{eqnarray*}
which is nothing but expression \eqref{eq:entrop_equi} and ends the proof. \endproof

\subsection{Onsager's symmetry}
\label{subsec_Onsager}

We note that the EME system \eqref{eq:hydro_limit} has a synthetic formulation in entropic variables, given by~\eqref{eq:EME_entrovar}. Now, we explore if the NSME system has a similar formulation. Before doing so, we need to introduce additional notations. 

For $\alpha, \beta \in \{1, \ldots, n+3\}$, we define an $n \times n$ matrix ${\mathbb X}^{\alpha \beta}$ as follows. 
\begin{eqnarray}
&& \hspace{-1cm}
{\mathbb X}^{\alpha \beta} = \varepsilon \mu \Theta \big( \mathrm{I}_n \delta^{\alpha \beta} + e_\beta \otimes e_\alpha - \frac{2}{n} e_\alpha \otimes e_\beta \big), \quad \forall \alpha, \, \beta = 1, \ldots, n, \label{eq:X1}\\
&& \hspace{-1cm}
{\mathbb X}^{\alpha \, n+1} = {\mathbb X}^{\alpha \, n+2} = ({\mathbb X}^{n+1 \, \alpha})^T = ({\mathbb X}^{n+2 \, \alpha})^T = \mathrm{0}_n, \quad \forall \alpha = 1, \ldots, n, \label{eq:X2}\\
&& \hspace{-1cm}
{\mathbb X}^{\alpha \, n+3} = ({\mathbb X}^{n+3 \, \alpha})^T = 2 \varepsilon \mu \Theta \big( u_\alpha \mathrm{I}_n  + u \otimes e_\alpha - \frac{2}{n} e_\alpha \otimes u \big), \quad \forall \alpha = 1, \ldots, n, \label{eq:X3}\\
&& \hspace{-1cm}
{\mathbb X}^{n+1 \, n+1} = \varepsilon \nu \mathrm{I}_n, \qquad
{\mathbb X}^{n+1 \, n+2} = ({\mathbb X}^{n+2 \, n+1})^T = {\mathbb X}^{n+2 \, n+2} = \mathrm{0}_n, \label{eq:X5}\\
&& \hspace{-1cm}
{\mathbb X}^{n+1 \, n+3} = ({\mathbb X}^{n+3 \, n+1})^T = (n+2) \varepsilon  \nu \Theta \mathrm{I}_n, \qquad
{\mathbb X}^{n+2 \, n+3} = ({\mathbb X}^{n+3 \, n+2})^T = \mathrm{0}_n, \label{eq:X7}\\
&& \hspace{-1cm}
{\mathbb X}^{n+3 \, n+3} = \varepsilon \Theta \Big[ \big( (n+2)^2 \nu \Theta + 4 \kappa \Theta + 4 \mu |u|^2 \big) \mathrm{I}_n + 4 \frac{n-2}{n} \mu \, u \otimes u \Big], \label{eq:X8}
\end{eqnarray}
where $\mathrm{0}_n$ is the $n \times n$ matrix with all entries equal to $0$ and where we recall that $(e_1, \ldots, e_n)$ denotes the canonical basis of ${\mathbb R}^n$. We note that ${\mathbb X}^{\alpha \beta} = ({\mathbb X}^{\beta \alpha})^T$ so that the $n(n+3) \times n(n+3)$ matrix ${\mathbb X}$ defined by blocks by ${\mathbb X} = ({\mathbb X}^{\alpha \beta})_{\alpha, \beta \in \{1, \ldots, n+3\}}$ is symmetric: 
\begin{equation} 
{\mathbb X} = {\mathbb X}^T. 
\label{eq:Onsag_sym}
\end{equation}
Now, we have the following 

\begin{proposition}
The NSME system \eqref{eq:diff_popu}-\eqref{eq:diff_energy} has the following equivalent formulation (for smooth solutions): 
\begin{equation}
\frac{\partial}{\partial t} \Big( \frac{\partial \Sigma}{\partial {\mathcal A}^\alpha} ({\mathcal A}) \Big) + \nabla_x \cdot \Big( \frac{\partial \Phi}{\partial {\mathcal A}^\alpha} ({\mathcal A}) \Big) = \nabla_x \cdot \Big(  \sum_{\beta = 1}^{n+3} {\mathbb X}^{\alpha \beta} \nabla_x {\mathcal A}^\beta \Big), \quad \forall \alpha = 1, \ldots, n+3. 
\label{eq:NSME_entrovar2}
\end{equation}
\label{prop:NSME_entrovar}
\end{proposition}

\begin{remark}
The form \eqref{eq:NSME_entrovar2} shows that the NSME system is consistent with the formalism of nonequilibrium thermodynamics \cite{de2013non}. Indeed, the contributions of diffusion to the time derivatives of the conservative variables ${\mathcal M}^\alpha = \frac{\partial \Sigma}{\partial {\mathcal A}^\alpha}$ are driven by divergence of fluxes which are linear combinations of the gradients of the entropic variables $\frac{\partial {\mathcal A}^\beta}{\partial {x_j}}$. This linear combination is described by the matrix ${\mathbb X}$. The fact that this matrix is symmetric reflects a general feature of nonequilibrium thermodynamics called \textbf{Onsager's symmetry}. When $\alpha \not = \beta$, Onsager's symmetry states that the contribution of the gradient of ${\mathcal A}^\beta$ to the time-derivative of ${\mathcal M}^\alpha$ is the adjoint of that of the gradient of ${\mathcal A}^\alpha$ to the time-derivative of ${\mathcal M}^\beta$. For instance, gradients of $C$ (generated by temperature gradients) induce temporal changes in the population equation (i.e. the equation for ${\mathcal N}$), as the non-zero value of ${\mathbb X}^{n+1 \, n+3}$ shows. But then, gradients in $A$ (generated by gradients in the population ${\mathcal N}$ for instance) induce temporal changes in the energy equation in the exact same amount (as ${\mathbb X}^{n+3 \, n+1} = {\mathbb X}^{n+1 \, n+3}$). Similar considerations hold true for any pair of components of the vector $D$, or betwen pairs $(D_\alpha, C)$ where $D_\alpha$ is any component of the vector $D$.   
\end{remark}

\noindent
\textbf{Proof.} As \eqref{eq:EME_entrovar} shows, the left-hand sides \eqref{eq:diff_popu}-\eqref{eq:diff_energy} can be written in the form of the left-hand side of \eqref{eq:NSME_entrovar2}. So, we are left to prove that the right-hand sides of  \eqref{eq:diff_popu}-\eqref{eq:diff_energy} can be written in the form of the right-hand side of \eqref{eq:NSME_entrovar2}. 

The right-hand sides of  Eqs. \eqref{eq:diff_mom}, \eqref{eq:diff_popu}, \eqref{eq:diff_mass} and \eqref{eq:diff_energy} correspond to the right-hand side of \eqref{eq:NSME_entrovar2} for $\alpha = 1, \ldots, n$, $\alpha = n+1$, $\alpha = n+2$ and $\alpha = n+3$ respectively. From the fact that the right-hand side of the mass conservation equation \eqref{eq:diff_mass} is identically $0$, we conclude that we can write it in the form  of the right-hand side of \eqref{eq:NSME_entrovar2} setting 
$$ {\mathbb X}^{n+2 \, \alpha} = \mathrm{0}_n, \qquad \forall \alpha = 1, \ldots, n=3. $$

Now, we consider the population conservation equation \eqref{eq:diff_popu}. Its right-hand side is equal to $\varepsilon \nabla_x \cdot (\nu \nabla_x \chi)$. Thanks to \eqref{eq:chi_def}, \eqref{eq:rel_rho_alpha} and \eqref{eq:map_entro2cons}, we can write 
$$ \chi = \log \frac{\rho}{Z(\beta,\Theta)} + \frac{n+2}{2} \log \Theta + \frac{n}{2} \log (2 \pi) = A + \frac{n+2}{2} \log \Big( - \frac{1}{2C} \Big) + \frac{n}{2} \log (2 \pi). $$
Hence, 
\begin{equation} 
\varepsilon \nu \nabla_x \chi  = \varepsilon \nu \big( \nabla_x A + (n+2) \Theta \nabla_x C \Big). 
\label{eq:eps_nu_nabla_chi}
\end{equation}
Comparing with \eqref{eq:NSME_entrovar2} for $\alpha = n+1$, we get
\begin{eqnarray*}
&& 
{\mathbb X}^{n+1 \, \alpha} = \mathrm{0}_n, \qquad \forall \alpha = 1, \ldots, n \quad \textrm{ and } \quad \alpha = n+2, \\
&&
{\mathbb X}^{n+1 \, n+1} = \varepsilon \nu \, \mathrm{I}_n, \qquad {\mathbb X}^{n+1 \, n+3} = (n+2) \varepsilon \nu \Theta \, \mathrm{I}_n. 
\end{eqnarray*}

Now, we turn ourselves to the momentum conservation equation \eqref{eq:diff_mom}. Using 
\eqref{eq:map_entro2cons}, we first observe that 
$$ 
\frac{\partial u_j}{\partial x_i} = - \frac{1}{2} \frac{\partial}{\partial x_i}  \Big( \frac{D_j}{C} \Big) = -\frac{1}{2C} \frac{\partial D_j}{\partial x_i} + \frac{D_j}{2 C^2} \frac{\partial C}{\partial x_i} = \Theta \frac{\partial D_j}{\partial x_i} + 2 \Theta u_j \frac{\partial C}{\partial x_i}. $$
Hence 
\begin{equation}
\sigma_{\alpha i}(u) = \Theta \Big[ \frac{\partial D_\alpha}{\partial x_i} + \frac{\partial D_i}{\partial x_\alpha} - \frac{2}{n} \sum_{j=1}^n \frac{\partial D_j}{\partial x_j} \delta_{\alpha i} \Big]  + 2 \Theta \Big[ u_i \frac{\partial C}{\partial x_\alpha} + u_\alpha \frac{\partial C}{\partial x_i} - \frac{2}{n} \sum_{j=1}^n u_j \frac{\partial C}{\partial x_j} \delta_{\alpha i} \Big] . 
\label{eq:stress_in_entrop_var}
\end{equation}
So, we have 
\begin{equation} 
\varepsilon \mu \sigma_{\alpha i} (u) = \textrm{\textcircled{a}} + \textrm{\textcircled{b}}, 
\label{eq:eps_mu_sigma}
\end{equation}
with 
\begin{eqnarray}
\textrm{\textcircled{a}} &=& \varepsilon \mu \Theta \Big[ \frac{\partial D_\alpha}{\partial x_i} + \frac{\partial D_i}{\partial x_\alpha} - \frac{2}{n} \sum_{j=1}^n \frac{\partial D_j}{\partial x_j} \delta_{\alpha i} \Big] \label{eq:sig_circle_a-1} \\
&=& \varepsilon \mu \Theta \sum_{j, \beta=1}^n \Big[ \delta_{ij} \delta_{\alpha \beta} + \delta_{j \alpha} \delta_{i \beta} - \frac{2}{n} \delta_{i \alpha} \delta_{j \beta} \Big] \frac{\partial D_\beta}{\partial x_j} \label{eq:sig_circle_a} \\
&=& \varepsilon \mu \Theta \sum_{j, \beta=1}^n \Big[ \mathrm{I}_n \delta_{\alpha \beta} + e_\beta \otimes e_\alpha - \frac{2}{n} e_\alpha \otimes e_\beta \Big]_{ij} \frac{\partial D_\beta}{\partial x_j}, \nonumber
\end{eqnarray}
and 
\begin{eqnarray}
\textrm{\textcircled{b}} &=& 2 \varepsilon \mu \Theta \Big[ u_i \frac{\partial C}{\partial x_\alpha} + u_\alpha \frac{\partial C}{\partial x_i} - \frac{2}{n} \sum_{j=1}^n u_j \frac{\partial C}{\partial x_j} \delta_{\alpha i} \Big] \label{eq:sig_circle_b-1} \\
&=& 2 \varepsilon \mu \Theta \sum_{j=1}^n \Big[ u_i \delta_{j \alpha} + u_\alpha \delta_{ij}  - \frac{2}{n} \delta_{i \alpha} u_j \Big] \frac{\partial C}{\partial x_j} \label{eq:sig_circle_b} \\
&=& 2 \varepsilon \mu \Theta \sum_{j=1}^n \Big[ u \otimes e_\alpha + u_\alpha \mathrm{I}_n  - \frac{2}{n} e_\alpha \otimes u \Big]_{ij} \frac{\partial C}{\partial x_j}, \nonumber
\end{eqnarray}
Comparing with \eqref{eq:NSME_entrovar2} for $\alpha = 1, \ldots, n$, it follows that 
\begin{eqnarray*}
{\mathbb X}^{\alpha \beta} &=& \varepsilon \mu \Theta  \Big[ \mathrm{I}_n \delta_{\alpha \beta} + e_\beta \otimes e_\alpha - \frac{2}{n} e_\alpha \otimes e_\beta \Big], \quad \forall \alpha, \, \beta \in \{1, \ldots, n \}, \\
{\mathbb X}^{\alpha \, n+1} &=& {\mathbb X}^{\alpha \, n+2} = \mathrm{0}_n, \quad \forall \alpha \in \{1, \ldots, n \}, \\ 
{\mathbb X}^{\alpha \, n+3} &=& 2 \varepsilon \mu \Theta  \Big[  u_\alpha \mathrm{I}_n +  u \otimes e_\alpha - \frac{2}{n} e_\alpha \otimes u \Big], \quad \forall \alpha \in \{1, \ldots, n \}.
\end{eqnarray*}

Finally, we consider the energy conservation equation \eqref{eq:diff_energy}. From \eqref{eq:eps_nu_nabla_chi}, we immediately get that 
\begin{equation} 
\varepsilon (n+2) \nu \Theta \nabla_x \chi  = (n+2) \varepsilon \nu \Theta \nabla_x A +  (n+2)^2 \varepsilon \nu \Theta^2 \nabla_x C.  
\label{eq:eps_nu_Theta_nachi}
\end{equation}
Then, using \eqref{eq:map_entro2cons}, we get
\begin{equation} 
2 \varepsilon \kappa \nabla_x \Theta = 4 \varepsilon \kappa \Theta^2 \nabla_x C. 
\label{eq:2eps_kap_naThet}
\end{equation}
Finally, thanks to \eqref{eq:eps_mu_sigma}, \eqref{eq:sig_circle_a} and \eqref{eq:sig_circle_b}, we have
$$ 2 \varepsilon \mu \big( \sigma(u) u \big)_i = \textrm{\textcircled{c}} + \textrm{\textcircled{d}}, $$
with 
\begin{eqnarray*}
\textrm{\textcircled{c}} &=& 2 \varepsilon \mu \Theta \sum_{j,k,\beta = 1}^n \Big[ \delta_{jk} \delta_{i \beta} + \delta_{j \beta} \delta_{ik} - \frac{2}{n} \delta_{ij} \delta_{\beta k} \Big] u_j \frac{\partial D_\beta}{\partial x_k} \\
&=& 2 \varepsilon \mu \Theta \sum_{j=1}^n \Big[ u_j \frac{\partial D_i}{\partial x_j} + u_j \frac{\partial D_j}{\partial x_i} - \frac{2}{n} u_i \frac{\partial D_j}{\partial x_j} \Big] \\
&=& 2 \varepsilon \mu \Theta \sum_{j, \beta = 1}^n \Big[u_j \delta_{i \beta} + u_\beta \delta_{ij} - \frac{2}{n} u_i \delta_{j \beta} \Big] \frac{\partial D_\beta}{\partial x_j} \\
&=& 2 \varepsilon \mu \Theta \sum_{j, \beta = 1}^n \Big[e_\beta \otimes u + u_\beta \mathrm{I}_n - \frac{2}{n} u \otimes e_\beta \Big]_{ij} \frac{\partial D_\beta}{\partial x_j}, 
\end{eqnarray*}
and 
\begin{eqnarray*}
\textrm{\textcircled{d}} &=& 4 \varepsilon \mu \Theta \sum_{j,k = 1}^n \Big[ u_i \delta_{jk} + u_j \delta_{ik} - \frac{2}{n} u_k \delta_{ij} \Big] u_j \frac{\partial C}{\partial x_k} \\
&=& 4 \varepsilon \mu \Theta \sum_{j=1}^n \Big[ \frac{n-2}{n} u_i u_j \frac{\partial C}{\partial x_j} + u_j^2 \frac{\partial C}{\partial x_i} \Big] \\
&=& 4 \varepsilon \mu \Theta \sum_{j = 1}^n \Big[\frac{n-2}{n} u_i u_j + |u|^2 \delta_{ij} \Big] \frac{\partial C}{\partial x_j} \\
&=& 4 \varepsilon \mu \Theta \sum_{j = 1}^n \Big[|u|^2 \mathrm{I}_n + \frac{n-2}{n} u \otimes u \Big]_{ij} \frac{\partial C}{\partial x_j}. 
\end{eqnarray*}
It follows that 
\begin{eqnarray} 
2 \varepsilon \mu \sigma(u) u &=& 2 \varepsilon \mu \Theta \Big\{ \sum_{\beta = 1}^n \Big[e_\beta \otimes u + u_\beta \mathrm{I}_n - \frac{2}{n} u \otimes e_\beta \Big] \nabla_x D_\beta \nonumber \\
&& \hspace{4cm} + 2 \Big[|u|^2 \mathrm{I}_n + \frac{n-2}{n} u \otimes u \Big] \nabla_x C \Big\}. \label{eq:2eps_mu_sigma_u}
\end{eqnarray}
Then, adding \eqref{eq:eps_nu_Theta_nachi}, \eqref{eq:2eps_kap_naThet} and \eqref{eq:2eps_mu_sigma_u}, we get 
\begin{eqnarray*}
{\mathbb X}^{n+3 \, \alpha} &=& 2 \varepsilon \mu \Theta \Big[ u_\alpha \mathrm{I}_n + e_\alpha \otimes u - \frac{2}{n} u \otimes e_\alpha \Big], \quad \forall \alpha = 1, \ldots, n, \\
{\mathbb X}^{n+3 \, n+1} &=& (n+2) \varepsilon \nu \Theta \mathrm{I}_n, \qquad  {\mathbb X}^{n+3 \, n+2} = \mathrm{0}_n, \\
{\mathbb X}^{n+3 \, n+3} &=& \varepsilon \Big[ (n+2)^2 \nu \Theta^2 + 4 \kappa \Theta^2 + 4 \mu \Theta |u|^2 \Big] \mathrm{I}_n + \frac{4(n-2)}{n} \varepsilon \mu \Theta u \otimes u. 
\end{eqnarray*}
Finally, we can check that all the formulas for ${\mathbb X}^{\alpha \beta}$ found in this proof coincide with Eqs.~\eqref{eq:X1}-\eqref{eq:X8}. \endproof

\subsection{Entropy dissipation}
\label{subsec_entrop_dissip}

Before stating the evolution equation for the entropy, we need the following 

\begin{lemma}[Nonnegativity of ${\mathbb X}$]~

\noindent
The matrix ${\mathbb X}$ is nonnegative. More precisely, let $(Y^\alpha)^{\alpha = 1, \ldots, n+3}$ be $n+3$ vectors of ${\mathbb R}^n$, of components $Y^\alpha = (Y^\alpha_i)_{i=1, \ldots, n}$. Denote also the $(i,j)$-the entry of the $n \times n$ matrix ${\mathbb X}^{\alpha \beta}$ by ${\mathbb X}^{\alpha \beta}_{ij}$. Then, we have 
\begin{eqnarray} 
&& \hspace{-1cm} 
\sum_{\alpha, \beta = 1}^{n+3} (Y^\alpha)^T {\mathbb X}^{\alpha \beta} Y^\beta = \varepsilon \nu \big| Y^{n+1} + (n+2) \Theta Y^{n+3} \big|^2 + 4 \varepsilon \kappa \Theta^2 \big| Y^{n+3} \big|^2 \nonumber \\
&& \hspace{0.5cm} 
+ \frac{\varepsilon \mu \Theta}{2} \sum_{i, \alpha=1}^n \bigg| \Big( Y^\alpha_i + Y^i_\alpha - \frac{2}{n} \sum_{j=1}^n Y^j_j \delta_{i \alpha} \Big) \nonumber \\
&& \hspace{3.cm} 
 + 2  \Big( u_i Y^{n+3}_\alpha + u_\alpha Y_i^{n+3} - \frac{2}{n} \sum_{j=1}^n u_j Y^{n+3}_j \delta_{i \alpha} \Big) \bigg|^2 , 
\label{eq:positivity_of_X}
\end{eqnarray}
and the right-hand side is nonnegative by \eqref{eq:nu_pos}. 
\label{lem:positivity_of_X}
\end{lemma}

\begin{remark}
The matrix ${\mathbb X}$ is not positive definite because \eqref{eq:positivity_of_X} offers no control on $Y^{n+2}$. It reflects the fact that ${\mathbb X}$ is singular because all lines and all columns associated with $Y^{n+2}$ have entries equal to zero. 
\end{remark} 

\noindent
\textbf{Proof.} Using \eqref{eq:X1} - \eqref{eq:X8} and the same computation as in \eqref{eq:sig_circle_a-1} and \eqref{eq:sig_circle_b-1}, we get 
\begin{eqnarray}
&& \hspace{-1cm} 
\sum_{\alpha, \beta = 1}^{n+3} (Y^\alpha)^T {\mathbb X}^{\alpha \beta} Y^\beta = \varepsilon \mu \Theta \Big\{ \sum_{i, \alpha = 1}^n Y_i^\alpha \Big[ Y_i^\alpha + Y_\alpha^i - \frac{2}{n} \sum_{j=1}^n Y_j^j \delta_{i \alpha} \Big] \nonumber \\
&& \hspace{3cm} 
+ 4 \sum_{i, \alpha = 1}^n Y_i^\alpha \Big[ u_i Y_\alpha^{n+3} + u_\alpha Y_i^{n+3} - \frac{2}{n} \sum_{j=1}^n u_j Y_j^{n+3} \delta_{i \alpha} \Big] \Big\} \nonumber \\
&& \hspace{-0.5cm} 
 + \varepsilon \nu |Y^{n+1}|^2 + 2 (n+2) \varepsilon \nu \Theta \, Y^{n+3} \cdot Y^{n+1} \nonumber \\
&& \hspace{-0.5cm}
+ \varepsilon \Big[ (n+2)^2 \nu \Theta^2 + 4 \kappa \Theta^2 + 4 \mu \Theta |u|^2 \Big] |Y^{n+3}|^2 + 4 \Big( 1 - \frac{2}{n} \Big) \varepsilon  \mu \Theta \, (u \cdot Y^{n+3})^2. \label{eq:YalXalbetYbet}
\end{eqnarray}
We have 
\begin{eqnarray*}
\sum_{i, \alpha = 1}^n Y_i^\alpha \Big[ Y_i^\alpha + Y_\alpha^i - \frac{2}{n} \sum_{j=1}^n Y_j^j \delta_{i \alpha} \Big] &=& \frac{1}{2} \sum_{i, \alpha = 1}^n \Big[ Y_i^\alpha + Y_\alpha^i \Big]  \Big[ Y_i^\alpha + Y_\alpha^i - \frac{2}{n} \sum_{j=1}^n Y_j^j \delta_{i \alpha} \Big]\\
&=& \frac{1}{2} \sum_{i, \alpha = 1}^n \Big[ Y_i^\alpha + Y_\alpha^i - \frac{2}{n} \sum_{j=1}^n Y_j^j \delta_{i \alpha} \Big]^2, 
\end{eqnarray*}
where the first equality is due to the fact that the tensor $(Y_i^\alpha + Y_\alpha^i - \frac{2}{n} \sum_{j=1}^n Y_j^j \delta_{i \alpha})_{i \alpha}$ is invariant by exchange of $i$ and $\alpha$ and the second equality come from the trace of this tensor being zero, i.e. 
$$ \sum_{i, \alpha=1}^n \Big[ Y_i^\alpha + Y_\alpha^i - \frac{2}{n} \sum_{j=1}^n Y_j^j \delta_{i \alpha} \Big] \delta_{i \alpha} = 0. $$
Thanks to the same algebra, we have 
\begin{eqnarray*} 
&& \hspace{-1cm} 
\sum_{i, \alpha = 1}^n Y_i^\alpha \Big[ u_i Y_\alpha^{n+3} + u_\alpha Y_i^{n+3} - \frac{2}{n} \sum_{j=1}^n u_j Y_j^{n+3} \delta_{i \alpha} \Big] \\
&& \hspace{0cm} 
= \frac{1}{2} \sum_{i, \alpha = 1}^n \Big[ Y_i^\alpha + Y_\alpha^i - \frac{2}{n} \sum_{j=1}^n Y_j^j \delta_{i \alpha} \Big] \Big[ u_i Y_\alpha^{n+3} + u_\alpha Y_i^{n+3} - \frac{2}{n} \sum_{j=1}^n u_j Y_j^{n+3} \delta_{i \alpha} \Big]. 
\end{eqnarray*}
and
\begin{eqnarray*}
&& \hspace{-1cm} 
\frac{2}{n} \sum_{i, \alpha = 1}^n \Big[ u_i Y_\alpha^{n+3} + u_\alpha Y_i^{n+3} - \frac{2}{n} \sum_{j=1}^n u_j Y_j^{n+3} \delta_{i \alpha} \Big] ^2 \\
&& \hspace{1cm} 
= \sum_{i, \alpha = 1}^n u_i Y_\alpha^{n+3} \Big[ u_i Y_\alpha^{n+3} + u_\alpha Y_i^{n+3} - \frac{2}{n} \sum_{j=1}^n u_j Y_j^{n+3} \delta_{i \alpha} \Big] \\
&& \hspace{1cm} 
= |u|^2 |Y^{n+3}|^2 + \Big( 1 - \frac{2}{n} \Big) (u \cdot Y^{n+3})^2. 
\end{eqnarray*}
Hence 
\begin{eqnarray*}
&& \hspace{-1cm} 
\varepsilon \mu \Theta \Big\{ \sum_{i, \alpha = 1}^n Y_i^\alpha \Big[ Y_i^\alpha + Y_\alpha^i - \frac{2}{n} \sum_{j=1}^n Y_j^j \delta_{i \alpha} \Big] \\
&& \hspace{4cm} 
+ 4 \sum_{i, \alpha = 1}^n Y_i^\alpha \Big[ u_i Y_\alpha^{n+3} + u_\alpha Y_i^{n+3} - \frac{2}{n} \sum_{j=1}^n u_j Y_j^{n+3} \delta_{i \alpha} \Big]  \\
&& \hspace{4cm} 
+ 4 |u|^2 |Y^{n+3}|^2 + 4 \Big( 1 - \frac{2}{n} \Big)  \, (u \cdot Y^{n+3})^2 \Big\}\\
&& \hspace{0.cm} 
= \frac{\varepsilon \mu \Theta}{2} \sum_{i, \alpha=1}^n \bigg| \Big( Y^\alpha_i + Y^i_\alpha - \frac{2}{n} \sum_{j=1}^n Y^j_j \delta_{i \alpha} \Big) \nonumber \\
&& \hspace{4cm} 
 + 2  \Big( u_i Y^{n+3}_\alpha + u_\alpha Y_i^{n+3} - \frac{2}{n} \sum_{j=1}^n u_j Y^{n+3}_j \delta_{i \alpha} \Big) \bigg|^2. 
\end{eqnarray*}
Inserting this into \eqref{eq:YalXalbetYbet}, we easily get \eqref{eq:positivity_of_X}. \endproof

We can now prove the 

\begin{proposition}[Entropy inequality for the NSME system]~

\noindent
Let $(\rho,u,\Theta,\beta)$ be a smooth solution of the NSME system. Let $S$ be the entropy (equivalently given by \eqref{eq:entrop_equi} or by \eqref{eq:legendre_transform}). Then, $S$ satisfies the following equation
\begin{eqnarray}
&&\hspace{-1cm}
\partial_t S + \nabla_x \cdot \tilde \phi = - \varepsilon \, \Big\{ \nu \, \big| \nabla_x \chi \big|^2  + \kappa \, \Big| \frac{\nabla_x \Theta}{\Theta} \Big|^2 + \frac{\mu}{2 \Theta} \sigma(u):\sigma(u) \, \Big\} \leq 0, 
\label{eq:entropy_NS}
\end{eqnarray}
where the entropy flux at the NSME level $\tilde \phi$ is given by  
\begin{eqnarray}
&&\hspace{-1cm}
\tilde \phi = \phi + \varepsilon \, \Big\{ - \nu \, \Big( \log \rho - \log Z - 1 - \frac{n}{2} \Big) \,  \nabla_x \chi +  \kappa \frac{\nabla_x \Theta}{\Theta} \Big\} , 
\label{eq:entropy_flux_NS}
\end{eqnarray}
with $\phi = S u$ being the kinetic entropy flux at equilibrium \eqref{eq:entrop_flux_equi}. The right-hand side of \eqref{eq:entropy_NS} is nonpositive because of \eqref{eq:nu_pos}. 
\label{prop:entropy_NS}
\end{proposition}

\begin{remark} 
Setting $\varepsilon = 0$ in \eqref{eq:entropy_flux_NS}, we get for smooth solutions of the EME system:  
$$\partial_t S + \nabla_x \cdot \phi = 0,$$
showing that $S$ and $\phi$ are the entropy and entropy flux pairs of the EME system (of course, for discontinuous solutions, the left-hand side is only smaller than or equal to $0$). 
\end{remark}

\noindent
\textbf{Proof.} Thanks to \eqref{eq:leg_transf_deriv}, we can write 
$$ \frac{\partial S}{\partial t} ({\mathcal M}) = \nabla_{{\mathcal M}} S({\mathcal M}) \bullet \frac{\partial {\mathcal M}}{\partial t} = {\mathcal A} \bullet \frac{\partial {\mathcal M}}{\partial t} = \sum_{\alpha = 1}^{n+3} {\mathcal A}^\alpha \frac{\partial {\mathcal M}^\alpha}{\partial t}. $$
Then, we use \eqref{eq:NSME_entrovar2} and get
$$  \frac{\partial S}{\partial t} ({\mathcal M}) + \sum_{\alpha = 1}^{n+3} {\mathcal A}^\alpha \nabla_x \cdot \Big( \frac{\partial \Phi}{\partial {\mathcal A}^\alpha} \Big) = \sum_{\alpha, \beta = 1}^{n+3} \nabla_x \cdot \Big( {\mathbb X}^{\alpha \beta} \nabla_x {\mathcal A}^\beta \Big) {\mathcal A}^\alpha. $$
On the one hand, we have
\begin{eqnarray*} \sum_{\alpha = 1}^{n+3} {\mathcal A}^\alpha \nabla_x \cdot \Big( \frac{\partial \Phi}{\partial {\mathcal A}^\alpha} \Big) &=& \sum_{\alpha = 1}^{n+3} \Big[ \nabla_x \cdot \Big( {\mathcal A}^\alpha \frac{\partial \Phi}{\partial {\mathcal A}^\alpha} \Big) -  \frac{\partial \Phi}{\partial {\mathcal A}^\alpha} \cdot \nabla_x {\mathcal A}^\alpha \Big]\\
&=& \nabla_x \cdot \Big( \sum_{\alpha = 1}^{n+3} {\mathcal A}^\alpha \frac{\partial \Phi}{\partial {\mathcal A}^\alpha} - \Phi({\mathcal A}) \Big). 
\end{eqnarray*}
On the other hand we can write 
$$  \sum_{\alpha, \beta = 1}^{n+3} \nabla_x \cdot \Big( {\mathbb X}^{\alpha \beta} \nabla_x {\mathcal A}^\beta \Big) {\mathcal A}^\alpha = \nabla_x \cdot \Big[ \sum_{\alpha, \beta = 1}^{n+3} {\mathbb X}^{\alpha \beta} \nabla_x {\mathcal A}^\beta {\mathcal A}^\alpha \Big] - 
\sum_{\alpha, \beta = 1}^{n+3}  (\nabla_x {\mathcal A}^\alpha)^T \, {\mathbb X}^{\alpha \beta} \, \nabla_x {\mathcal A}^\beta . $$
Thus, we can write 
\begin{equation} \partial_t S + \nabla_x \cdot \tilde \phi = - 
\sum_{\alpha, \beta = 1}^{n+3}  (\nabla_x {\mathcal A}^\alpha)^T \, {\mathbb X}^{\alpha \beta} \, \nabla_x {\mathcal A}^\beta, 
\label{eq:entrop_abstract}
\end{equation}
with 
\begin{equation} 
\tilde \phi = \sum_{\alpha = 1}^{n+3} {\mathcal A}^\alpha \frac{\partial \Phi}{\partial {\mathcal A}^\alpha} - \Phi({\mathcal A}) - \sum_{\alpha, \beta = 1}^{n+3} {\mathbb X}^{\alpha \beta} \nabla_x {\mathcal A}^\beta {\mathcal A}^\alpha. 
\label{eq:entrop_dissip_abstract}
\end{equation}

Now, with \eqref{eq:positivity_of_X} applied with $Y^\alpha = \nabla_x {\mathcal A}^\alpha$ and with \eqref{eq:eps_nu_nabla_chi}, \eqref{eq:stress_in_entrop_var}, \eqref{eq:2eps_kap_naThet}, we have 
\begin{eqnarray}
\sum_{\alpha, \beta = 1}^{n+3}  (\nabla_x {\mathcal A}^\alpha)^T \, {\mathbb X}^{\alpha \beta} \, \nabla_x {\mathcal A}^\beta &=& \varepsilon \nu \big|\nabla_x A + (n+2) \Theta \nabla_x C \big|^2 + 4 \varepsilon \kappa \Theta^2 \big|\nabla_x C \big|^2 \nonumber \\
&& 
+ \frac{\varepsilon \mu \Theta}{2} \sum_{i, \alpha=1}^n \bigg| \Big( \frac{\partial D_\alpha}{\partial x_i} + \frac{\partial D_i}{\partial x_\alpha} - \frac{2}{n} \sum_{j=1}^n \frac{\partial D_j}{\partial x_j} \delta_{i \alpha} \Big) \nonumber \\
&& 
 + 2  \Big( u_i \frac{\partial C}{\partial x_\alpha} + u_\alpha \frac{\partial C}{\partial x_i} - \frac{2}{n} \sum_{j=1}^n u_j \frac{\partial C}{\partial x_j} \delta_{i \alpha} \Big) \bigg|^2 \nonumber \\
&=& \varepsilon \nu \big| \nabla_x \chi \big|^2 + \varepsilon \kappa \Big| \frac{\nabla_x \Theta}{\Theta} \Big|^2 + \frac{\varepsilon \mu}{2 \Theta} \sigma(u):\sigma(u),  
\label{eq:dissip_part}
\end{eqnarray}
which, inserted into \eqref{eq:entrop_abstract}, yields \eqref{eq:entropy_NS}. 

Now, we compute $\tilde \phi$ given by \eqref{eq:entrop_dissip_abstract}. Using \eqref{eq:Big_Phi_def}, \eqref{eq:entrop_flux_equi} and \eqref{eq:entrop_flux_alt}, we see that 
\begin{eqnarray} \sum_{\alpha = 1}^{n+3} \frac{\partial \Phi}{\partial {\mathcal A}^\alpha} {\mathcal A}^\alpha - \Phi &=& \sum_{m \geq 1} \int_{{\mathbb R}^n} \frac{m^{\frac{n}{2}}}{\gamma_m} \, e^{\mu_m \bullet {\mathcal A}} \, \big( \mu_m \bullet {\mathcal A} - 1 \big) \, v \, dv \nonumber\\
&=& \sum_{m \geq 1} \int_{{\mathbb R}^n} \rho (M_{u,\Theta,\beta})_m \, \Big( \log \frac{\gamma_m \rho (M_{u,\Theta,\beta})_m}{m^{\frac{n}{2}}} - 1 \Big) \, v \, dv  \nonumber \\
&=& \phi (\rho M_{u,\Theta,\beta}) = \phi = uS. \label{eq:express_phi_alternate}
\end{eqnarray}
Then, because $\sum_{\beta = 1}^{n+3} {\mathbb X}^{\alpha \beta} \nabla_x {\mathcal A}^\beta$ is another way to write the right-hand sides of the NSME system \eqref{eq:diff_popu}-\eqref{eq:diff_energy}, we have 
\begin{eqnarray}
\sum_{\alpha, \beta = 1}^{n+3} {\mathbb X}^{\alpha \beta} \nabla_x {\mathcal A}^\beta {\mathcal A}^\alpha &=& \varepsilon \Big[ \mu \sigma(u) D + \nu A \nabla_x \chi + C \big( (n+2) \nu \Theta \nabla_x \chi + 2 \kappa \nabla_x \Theta + 2 \mu \sigma(u) u \big) \Big] \nonumber \\
&=& \varepsilon \nu \nabla_x \chi \Big( \log \rho - \log Z - \frac{n+2}{2} \Big) - \kappa \frac{\nabla_x \Theta}{\Theta}. \label{eq:express_tilphi_alternate}
\end{eqnarray}
where the second equality comes from \eqref{eq:map_entro2cons}, \eqref{eq:rel_rho_alpha}. Now, inserting \eqref{eq:express_phi_alternate} and \eqref{eq:express_tilphi_alternate} into \eqref{eq:entrop_dissip_abstract} leads to \eqref{eq:entropy_flux_NS} and ends the proof. \endproof

\begin{remark}
It is possible to derive the entropy / entropy-dissipation identities \eqref{eq:entropy_NS}, \eqref{eq:entropy_flux_NS} directly from the NSME system \eqref{eq:diff_popu}-\eqref{eq:diff_energy}. However, our proof reveals the structure \eqref{eq:entrop_abstract}, \eqref{eq:entrop_dissip_abstract} of these identities. This structure is generic to all systems deriving from thermodynamic principles. 
\end{remark}

From Prop. \ref{prop:entropy_NS}, we deduce the following corollary, whose proof is immediate. It shows that the NSME system is compatible with the second law of thermodynamics. 

\begin{corollary}[Entropy decay in the NSME system]~

\noindent
Let $(\rho,u,\Theta,\beta)$ be a smooth solution of the NSME system in a smooth domain $\Omega$ such that the normal entropy flux $\tilde \phi \cdot {\mathbf n} = 0$ across the boundary $\partial \Omega$ vanishes (where ${\mathbf n}$ is the outward unit normal to $\partial \Omega$). Then, the integral of $S$ over $\Omega$ is non-increasing in time, i.e. 
$$ t_1 < t_2 \quad \Longleftrightarrow \quad \int_{\Omega} S(x,t_1) \, dx \geq \int_{\Omega} S(x,t_2) \, dx . $$
Furthermore, if $\sigma(u)$, $\nabla_x \Theta$ or $\nabla_x \chi$ are non-zero over a non-negligible subset of $\Omega$ for all time within an open subinterval of $(t_1,t_2)$, then, the previous inequality is strict.  
\label{cor:entropy_decay}
\end{corollary}

The entropy dissipation inequality has a mathematical consequence which we highlight on the linearized NSME system about a uniform state described by the entropic variable~${\mathcal A}_0$. By \eqref{eq:NSME_entrovar2}, this linearized system can be written
\begin{eqnarray}
&& \hspace{-1cm}
\sum_{\beta = 1}^{n+3} \frac{\partial^2 \Sigma}{\partial {\mathcal A}^\alpha \partial {\mathcal A}^\beta}({\mathcal A}_0)  \frac{\partial {\mathcal A}^\beta}{\partial t} + \sum_{\beta = 1}^{n+3} \frac{\partial^2 \Phi}{\partial {\mathcal A}^\alpha  \partial {\mathcal A}^\beta}({\mathcal A}_0) \cdot \nabla_x {\mathcal A}^\beta \nonumber \\
&& \hspace{3cm}
 = \nabla_x \cdot \Big(  \sum_{\beta = 1}^{n+3} {\mathbb X}^{\alpha \beta}({\mathcal A}_0) \nabla_x {\mathcal A}^\beta \Big), \quad \forall \alpha = 1, \ldots, n+3. 
\label{eq:NSME_linear}
\end{eqnarray}
For this system, we have the 

\begin{proposition}[Parabolicity of the linearized NSME system]~

\noindent
The linearized NSME system \eqref{eq:NSME_linear} is parabolic. In particular, consider it on a domain~$\Omega$ with Dirichlet boundary conditions ${\mathcal A}|_{\partial \Omega} = 0$. Then, we have the following energy identity:  
\begin{equation}
\frac{1}{2}  \frac{d}{dt} \int_{\Omega} \big( (\nabla_{{\mathcal A}}^2 \Sigma)({\mathcal A}_0) {\mathcal A} \big) \bullet {\mathcal A} \, dx = \int_{\Omega} \sum_{\alpha, \beta = 1}^{n+3} (\nabla_x {\mathcal A}^\beta)^T {\mathbb X}_{\alpha \beta}({\mathcal A}_0) \nabla_x {\mathcal A}^\alpha \, dx \leq 0. 
\label{eq:energy_identity}
\end{equation}
There exists a constant $C \geq 1$ such that for any $0 \leq t_1 < t_2$, we have 
\begin{equation}
\int_{\Omega} ({\mathcal A} \bullet {\mathcal A}) (x,t_2) \, dx \leq C \int_{\Omega} ({\mathcal A} \bullet {\mathcal A}) (x,t_1) \, dx. 
\label{eq:norm_identity}
\end{equation}
\label{prop:parabolicity}
\end{proposition}

\noindent
\textbf{Proof.} Since the matrix $(\nabla_{{\mathcal A}}^2 \Sigma)({\mathcal A}_0)$ is positive definite, parabolicity is equivalent to the positivity of the matrix ${\mathbb X}$ which was proved in Lemma \ref{lem:positivity_of_X}. 

We can re-write~\eqref{eq:NSME_linear} as 
\begin{eqnarray*}
&& \hspace{-1cm}
(\nabla_{{\mathcal A}}^2 \Sigma)({\mathcal A}_0) \frac{\partial {\mathcal A}}{\partial t} + \sum_{i = 1}^{n} (\nabla_{{\mathcal A}}^2 \Phi_i)({\mathcal A}_0) \frac{\partial {\mathcal A}}{\partial x_i}  = \nabla_x \cdot \Big(  \sum_{\beta = 1}^{n+3} {\mathbb X}^{\alpha \beta}({\mathcal A}_0) \nabla_x {\mathcal A}^\beta \Big). 
\end{eqnarray*}
Then, we multiply this equation by ${\mathcal A}^\alpha$ componentwise, sum them over $\alpha$ and integrate the result over the domain $\Omega$. Owing to the fact that $(\nabla_{{\mathcal A}}^2 \Sigma)({\mathcal A}_0)$ is symmetric, the first term gives the first term of \eqref{eq:energy_identity}. Because the matrices $(\nabla_{{\mathcal A}}^2 \Phi_i)({\mathcal A}_0)$ (for $i=1, \ldots, n$) are also symmetric, the second term leads to 
\begin{eqnarray*} 
&&
\frac{1}{2} \sum_{i = 1}^{n} \int_{\Omega} \frac{\partial}{\partial x_i} \Big[ \big( (\nabla_{{\mathcal A}}^2 \Phi_i)({\mathcal A}_0) {\mathcal A} \big) \bullet {\mathcal A} \Big] \, dx  \\
&&\hspace{1cm} 
= \frac{1}{2} \sum_{i = 1}^{n} \int_{\partial \Omega}  \big( (\nabla_{{\mathcal A}}^2 \Phi_i)({\mathcal A}_0) {\mathcal A} \big) \bullet {\mathcal A} \, {\mathbf n}_i\, d\gamma(x) = 0, \end{eqnarray*}
due to the boundary conditions. Here $({\mathbf n}_i)_{i=1}^n$ are the $n$ components of the outward unit normal ${\mathbf n}$ to the boundary $\partial \Omega$ and $d\gamma(x)$ is the surface measure on $\partial \Omega$. The computation of the last term is done in a similar way and this leads to \eqref{eq:energy_identity}. 

Finally, since the matrix $(\nabla_{{\mathcal A}}^2 \Sigma)({\mathcal A}_0)$ is positive definite, there exist two constants $0 < C_1 \leq C_2$ such that 
$$ C_1  {\mathcal A} \bullet {\mathcal A} \, \leq \, \big( (\nabla_{{\mathcal A}}^2 \Sigma)({\mathcal A}_0) {\mathcal A} \big) \bullet {\mathcal A} \, \leq \, C_2 {\mathcal A} \bullet {\mathcal A}. $$
Then, \eqref{eq:norm_identity} follows with $C = C_2/C_1$. \endproof

\setcounter{equation}{0}
\section{Conclusion}
\label{sec_conclu}

In this work, we have derived a new Boltzmann operator for binary collisions with mass-exchange. We have investigated its properties, notably entropy dissipation and equilibria, and derived macroscopic models of Euler or Navier-Stokes types in the hydrodynamic regime. Finally, we have shown that the Navier-Stokes type macroscopic model complies with the requirements of nonequilibrium thermodynamics, namely, Onsager's reciprocity and entropy dissipation. This work opens many interesting research areas. First, existence and uniqueness of solutions to this Boltzmann equation or its Navier-Stokes counterpart remain to be proved even in the spatially homogeneous case (see \cite{cercignani2013mathematical,ukai2006mathematical,villani2002review} for reviews on these questions in the case of the classical Boltzmann equation and \cite{bresch2018global,feireisl2004dynamics,lions1996mathematical} in the case of the compressible Navier-Stokes equation). Then, conducting a  rigorous study of the hydrodynamic limit in the spirit of Caflisch's seminal work \cite{caflisch1980fluid} would be of great interest. Other kinds of hydrodynamic limits could be investigated, such as those leading to incompressible systems \cite{bardos1991fluid,bardos1993fluid,de1989incompressible,saint2009hydrodynamic}. This study has been partly motivated by the dynamics of animal groups. So, pursuing in this direction, one could replace the rarefied gas dynamics Boltzmann collision operator by an operator modelling collective dynamics such as the Vicsek-Fokker-Planck model \cite{briant2022cauchy,degond2015phase,degond2008continuum,figalli2018global,gamba2016global} or the Bertin-Droz-Gr\'egoire collision model \cite{bertin2006boltzmann,bertin2009hydrodynamic}. However, in the latter case, the study would be made difficult by the non-availability of analytic formulas for the equilibria \cite{carlen2015boltzmann}.

\appendix

\vspace{0.8cm}
\noindent
\textbf{\Large Appendix}

\vspace{-0.3cm}
\setcounter{section}{0}
\section{Remarkable formulas}
\label{app_remark_form}

In this appendix, we recall the following formulas whose proofs are classical. 

\begin{eqnarray} 
&&\hspace{-1.5cm}
\int_{{\mathbb R}^n} e^{- \frac{m |v-u|^2}{2 \Theta}} \, |v-u|^{2p} \, dv = \Big( \frac{2 \pi \Theta}{m} \Big)^{\frac{n}{2}}  \, \prod_{k=0}^{p-1} (n+2k) \, \Big( \frac{\Theta}{m} \Big)^p,
\label{eq:moments_maxwellian_1}\\
&&\hspace{-1cm}
\mbox{} \nonumber \\
&&\hspace{-1.5cm}
\int_{{\mathbb R}^n} e^{- \frac{m |v-u|^2}{2 \Theta}} \, |v-u|^{2p} \, (v-u) \otimes (v-u) dv =  \Big( \frac{2 \pi \Theta}{m} \Big)^{\frac{n}{2}} \,   \prod_{k=1}^p (n+2k) \, \Big( \frac{\Theta}{m} \Big)^{p+1} \, \mathrm{I}_n, 
\label{eq:moments_maxwellian_2} \\
&&\hspace{-1cm}
\mbox{} \nonumber \\
&&\hspace{-1.5cm}
\int_{{\mathbb R}^n} e^{- \frac{m |v-u|^2}{2 \Theta}} \, (v-u)^{\otimes 4} dv =  \Big( \frac{2 \pi \Theta}{m} \Big)^{\frac{n}{2}} \, \Big( \frac{\Theta}{m} \Big)^2 \, {\mathbb E} \label{eq:moments_maxwellian_3}
\end{eqnarray}
where ${\mathbb E}$ is the four-rank tensor given by 
$$ 
{\mathbb E}_{i j k \ell} = \delta_{ij} \delta_{k \ell} + \delta_{ik} \delta_{j \ell} + \delta_{i \ell} \delta_{j k}. 
$$

\section*{Declarations}

\textbf{Competing interests.} The authors declare no competing interests. 

\noindent
\textbf{Funding.} No funds, grants, or other support was received.

\noindent
\textbf{Data statement.} No new data has been produced in the course of this research.


\end{document}